\documentclass[final,5p,authoryear,times]{elsarticle}
\pdfoutput=1
\usepackage[breaklinks,colorlinks,linkcolor=blue,citecolor=blue,urlcolor=blue]{hyperref}
\usepackage{graphicx}
\usepackage{amsfonts,amsmath,amssymb}
\usepackage[italic]{mathastext}
\usepackage[utf8]{inputenc}
\usepackage{textcomp}
\usepackage{caption}
\usepackage{flushend}

\usepackage[ruled]{algorithm}
\usepackage[noend]{algpseudocode}
\alglanguage{pseudocode}

\newcommand{\red}[1]{#1}

\newcommand{\x}{\mathbf{x}}
\newcommand{\y}{\mathbf{y}}
\newcommand{\z}{\mathbf{z}}
\newcommand{\bb}{\mathbf{b}}
\newcommand{\tens}[1]{\mathsf{#1}}
\newcommand{\X}{\tens{X}}
\newcommand{\Y}{\tens{Y}}
\newcommand{\bB}{\tens{B}}
\newcommand{\bS}{\tens{S}}
\newcommand{\bT}{\tens{T}}
\newcommand{\balpha}{\boldsymbol{\alpha}}
\newcommand{\btheta}{\boldsymbol{\theta}}
\newcommand{\bmu}{\boldsymbol{\mu}}
\newcommand{\bSigma}{\tens{\Sigma}}
\newcommand{\like}{\mathcal{L}}
\newcommand{\loglike}{\ln\mathcal{L}}

\newcommand{\python}{{\sc python}}
\newcommand{\algo}{{\tt GMMis}}
\newcommand{\repo}{https://github.com/pmelchior/pyGMMis}
\newcommand{\chandra}{\emph{Chandra}}

\DeclareRobustCommand{\object}[1]{%
   #1%
}

\def\apj{ApJ}%
\def\apjs{ApJS}%
\ifx \undefined \rsquo     \def \rsquo {'} \fi

\journal{Astronomy \& Computing}
\begin{document}
\begin{frontmatter}

\title{Filling the gaps: Gaussian mixture models from noisy, truncated or incomplete samples}

\author[1]{Peter Melchior}
\ead{peter.melchior@princeton.edu}
\author[1]{Andy D. Goulding}
\address[1]{Department of Astrophysical Sciences, Princeton University, Princeton, NJ 08544, USA}

\begin{abstract}
Astronomical data often suffer from noise and incompleteness.
We extend the common mixtures-of-Gaussians density estimation approach to account for situations with a known sample incompleteness by simultaneous imputation from the current model.
The method, called \algo, generalizes existing Ex\-pec\-tation-Maximization techniques for truncated data to arbitrary truncation geometries and probabilistic rejection processes, as long as they can be specified and do not depend on the density itself.
The method accounts for independent multivariate normal measurement errors for each of the observed samples and recovers an estimate of the error-free distribution from which both observed and unobserved samples are drawn.
It can perform a separation of a mixtures-of-Gaussian signal from a specified background distribution whose amplitude may be unknown.
We compare \algo\ to the standard Gaussian mixture model for simple test cases with different types of incompleteness,  and apply it to observational data from the NASA \chandra\ X-ray telescope.
The \python\ code is released as an open-source package at \url{\repo}.
\end{abstract}


\end{frontmatter}

\section{Introduction}

The Gaussian mixture model (GMM) is an important tool for many data analysis tasks, usually employed to approximate a potentially complex density distribution for which there is no suitable parametric form, or to perform a clustering analysis. 
Its importance is reflected in the wide range of applications and the number of extensions it has received \citep[e.g.][]{McLachlan00.1, Mengersen11.1}. 
In this work, we will employ the Expectation-Maximization (EM) algorithm to account for ``incomplete'' samples, i.e. a generalization of truncated samples where data from arbitrarily shaped regions of the feature space have a specified probability of not being reported. Also known as ``sample selection bias'', it constitutes a long-standing problem for robust inference of population statistics in many scientific disciplines.

In observational astronomy, sample incompleteness occurs frequently, caused e.g. by gaps between sensors or by proximity to bright objects that render a portion of the observation useless.
As a milder form of incomplete data, long-running surveys routinely encounter variations in how well the sky can be observed, for instance when the transparency of the atmosphere or the brightness of the moon changes.
As a consequence, surveys exhibit complicated completeness functions for the density of observed samples and derived data products \citep[e.g.][]{Leistedt16.1}.

Several approaches for dealing with truncated data within the context of Gaussian mixtures have been presented. Given a simple truncation boundary, one can analytically integrate the Gaussian density distribution over the unobserved regions \citep[e.g.][and references therein]{Wolynetz79.1,Lee12.1}. 
\citet{McLachlan88.1} and \citet{Cadez02.1} developed a method for binned data, which requires integration of the Gaussian distribution within bins, and realized that any truncated data can be thought to belong to one extra bin, for which the moments of the GMM are already fully specified by the values in the observed bins. 
Provided the binning is sufficiently fine, this approach allows for arbitrarily shaped truncation boundaries at the cost of introducing, and integrating over, bins for the entire observed region.

Our approach instead follows a proposal by \citet{Dempster77.1} that an estimate of the missing data be drawn from the current model and the EM be run on the combined (observed and estimated missing) data until convergence.
This approach does not require binning and provides flexibility and computational efficiency.
The method we develop here is similar to Multiple Imputation \citep{Rubin87.1} and Stochastic EM \citep{Diebolt1996} schemes for samples with missing features, and we will draw from the terminology developed in that context to clarify what kind of incompleteness our density-estimation method can account for.
While inspired by these earlier works, our contribution enables the treatment of missing samples, not just missing features, an extension that renders the GMM applicable to a wide range of observational cases.
We also incorporate the "Extreme Deconvolution" approach of \citet{Bovy11.1} to account for errors of the observed samples and clarify the interplay between noise and missingness.

The outline of the paper is as follows: In \autoref{sec:theory} we describe the EM algorithm for GMM optimization, show how it can accommodate noisy samples, and generalize it for incomplete and potentially noisy samples.
We discuss several practical extensions of the algorithm in \autoref{sec:practice} and demonstrate the performance of the proposed algorithm on a variety of test cases in \autoref{sec:experiments}.
We present an example application with data from the NASA \chandra\ X-ray telescope, exhibiting spatially varying completeness and positional errors, in \autoref{sec:application} and conclude in \autoref{sec:conclusions}.

The method is implemented in pure \python, scalable to millions of samples and thousands of GMM components, and publicly released at \url{\repo}.

\section{GMM from noisy, incomplete samples}
\label{sec:theory}

For a general mixture model over a $d$-dimensional feature space $\mathbb{R}^d$, the probability density function (PDF) is
\begin{equation}
\label{eq:gmm}
p(\x \mid \btheta) = \red{\frac{1}{Z(\btheta)}}\sum_{k=1}^K \alpha_k\, p(\x \mid \btheta_k)\mathrm{,}
\end{equation}
with mixing weights that obey $\sum_k\alpha_k=1$. The component density functions $p(\x \mid \btheta_k)$ with parameters $\btheta_k$, the list of which $\lbrace\btheta_1,\btheta_2,\dots\rbrace$ we simply write as $\btheta$, are assumed to be normalized, \red{so that the overall normalization constant $Z(\btheta)=1$ and can be dropped.} We will, for the sake of brevity, abbreviate $p(\x \mid \btheta_k)$ as $p_k(\x)$. 
For a GMM, $p_k$ is given by a multivariate normal distribution function,
\begin{equation}
p_k(\x) = \mathcal{N}(\x \mid \bmu_k,\bSigma_k) \equiv \frac{\exp\big[-\frac{1}{2}(\x-\bmu_k)^\top\bSigma_k^{-1}(\x-\bmu_k)\big]}{\sqrt{(2\pi)^d \det(\bSigma_k)}}\mathrm{,}
\end{equation}
the complete set of parameters per component is thus $\alpha_k$, $\bmu_k$, and $\bSigma_k$.
The corresponding mixture log-likelihood for noise-free observations $\mathcal{D}=\lbrace\x_i\rbrace$ with $i\in\lbrace1,\dots,N\rbrace$ given fixed parameters is
\begin{equation}
\label{eq:logL}
\loglike(\mathcal{D}\mid\btheta) \equiv \sum_i^N \ln \sum_k^K \alpha_k p_k(\x_i).
\end{equation}
This form is also called \emph{uncategorized} log-likelihood \citep{Titterington85.1} because there is no information about which component $k$ generated sample $i$.
We introduce the discrete indicator $\mathbf{S} = (S_1,\dots,S_N)$ with $S_i = k$ if (and only if) $\x_i$ is generated by component $k$. 
By rewriting $\alpha_k p_k(\x_i) = q_{ik} p(\x_i \mid \balpha,\bmu,\bSigma)$ with the conditional probability $q_{ik}$ that $\x_i$ is generated by component $k$, i.e. $q_{ik}\equiv \text{Pr}(S_i=k \mid \x_i,\balpha,\bmu,\bSigma)$,  we can form the 
\emph{complete-data} log-likelihood
\begin{equation}
\label{eq:complete-logL}
\begin{split}
p(\x,\mathbf{S} \mid \balpha,\bmu,\bSigma)=\ &p(\x \mid \balpha,\bmu,\bSigma)\ + \\
&\sum_i^N\sum_k^K I_{S_i=k} \ln\text{Pr}(S_i=k \mid \x_i,\balpha,\bmu,\bSigma),
\end{split}
\end{equation}
the sampling distribution of the complete data $(\x,\mathbf{S})$ given the parameters of the model $(\balpha,\bmu,\bSigma)$.
The second term of the RHS describes the additional information from knowing the allocation $\mathbf{S}$ of samples to components \citep{Fruehwirth06}.

\subsection{Standard EM algorithm}
\label{sec:standard-EM}

As the indicator $\mathbf{S}$ is not known or directly observable, we need to estimate it by classifying each observation.
With Bayes' rule, and dropping the conditional dependence on the full set of model parameters $(\balpha,\bmu,\bSigma)$ in all terms, the classification is given by
\begin{equation}
\text{Pr}(S_i=k \mid \x_i) = \frac{\text{Pr}(\mathbf{X} = \x_i \mid S_i=k)\,\text{Pr}(S_i=k)}{\sum_j^K\text{Pr}(\mathbf{X} = \x_i \mid S_i=j)\,\text{Pr}(S_i=j)}.
\end{equation}
With it one can re-estimate the parameters of the model from the weighted moments of $\x$ given the estimate of $\mathbf{S}$.
That is the central idea of the EM procedure, which first estimates the classification (the E-step) and then updates the model parameters (the M-step).
Assuming flat prior distributions $\text{Pr}(S_i=k)$ and using the definition of $q_{ik}$ from above, we get
\begin{equation}
\label{eq:EM1}
\begin{split}
\text{E-step:}\ \ q_{ik} &\leftarrow \frac{\alpha_k p_k(\x_i)}{\sum_j \alpha_j p_j(\x_i)}\\
\text{M-step:}\ \ \alpha_k &\leftarrow \frac{1}{N} \sum_i q_{ik} \equiv \frac{1}{N} q_k\\
 \bmu_k & \leftarrow \frac{1}{q_k} \sum_i q_{ik}\x_i\\
 \bSigma_k & \leftarrow \frac{1}{q_k} \sum_i q_{ik}\left[ (\bmu_k - \x_i)(\bmu_k - \x_i)^\top\right].
\end{split}
\end{equation}
\citet{Dempster77.1}, later corrected by \citet{Wu83.1}, showed that repeated iterations of these two steps monotonically converge to a local maximum of $\like$.

\subsection{EM algorithm with noisy samples}
\label{sec:EM-noisy}

Observed data often exhibit measurement uncertainties, which we assume to be additive and Gaussian:
\begin{equation}
\label{eq:noisy_data}
\y_i \equiv \x_i + \mathbf{e}_i\mathrm{,}
\end{equation}
where $\mathbf{e}_i \sim \mathcal{N}(\mathbf{0},\, \bS_i)$.
We want to emphasize that we do not assume that errors are identically distributed, only that they are independent, Gaussian, and that their covariance matrices are known.
\citet{Bovy11.1} derived an extended EM algorithm that maximizes the likelihood of the noise-free density $p(\x)$ from noisy samples $\y_i$.
The key insight is that one can marginalize over the unknown values $\x_i$ and still obtain a GMM. 
With $\bT_{ik} \equiv \bSigma_k + \bS_i$ the EM procedure amounts to 
\begin{equation}
\label{eq:EM2}
\begin{split}
\text{E-step:}\ \ q_{ik} &\leftarrow \frac{\alpha_k p_k(\y_i \mid \bmu_k,\bT_{ik})}{\sum_j \alpha_j p_j(\y_i \mid \bmu_j,\bT_{ij})}\\
\bb_{ik} & \leftarrow \bmu_k + \bSigma_k\bT_{ik}^{-1}(\y_i-\bmu_k)\\
\bB_{ik} & \leftarrow \bSigma_k - \bSigma_k\bT_{ik}^{-1}\bSigma_k\\
\text{M-step:}\ \ \alpha_k &\leftarrow \frac{1}{N} \sum_i q_{ik} \equiv \frac{1}{N}q_k\\
 \bmu_k & \leftarrow \frac{1}{q_k} \sum_i q_{ik}\bb_{ik}\\
 \bSigma_k & \leftarrow \frac{1}{q_k} \sum_i q_{ik}\left[ (\bmu_k - \bb_{ik})(\bmu_k - \bb_{ik})^\top + \bB_{ik}\right],
\end{split}
\end{equation}
By including measurement uncertainties in the definition of $q_{ik}$, the M-step is almost unchanged: the role of $\x_i$, which is not directly observable, is played by $\bb_{ik}$, and the covariances get extra contributions $\bB_{ik}$.
All of those modifications vanish when $\bS_i \rightarrow \tens{0}$.
The resulting EM algorithm still converges monotonically to a local maximum of $\like$.

\subsection{EM algorithm with incomplete samples}
\label{sec:gmmis}

We will now account for sample incompleteness, by which we mean the following: 
the probability of a sample $\x$ being observed at all is determined by a completeness function, $\Omega(\x):\ \mathbb{R}^d\rightarrow[0,1]$, which means that in any region $\mathcal{R}\subset\mathbb{R}^d$ the density of observed samples can systematically deviate from the true density in that region. 
\red{Formally,
\begin{equation}
\label{eq:observed_gmm}
p_o(\x \mid \btheta, \Omega) = \frac{1}{Z(\btheta, \Omega)} \Omega(\x)\, p(\x \mid \btheta)\mathrm{,}
\end{equation}
with
\begin{equation}
\label{eq:observed_Z}
Z(\btheta, \Omega) \equiv \int \mathrm{d}\x\ \Omega(\x)\, p(\x \mid \btheta).
\end{equation}
The corresponding log-likelihood is
\begin{equation}
\label{eq:obs-likelihood}
\loglike_o(\mathcal{D}\mid \btheta, \Omega) = \sum_i^N \ln p_o(\x_i \mid \btheta, \Omega).
\end{equation}
}
Such a situation may arise if a measurement device is incapable of recording samples from $\mathcal{R}$, so that $\Omega(\x)=0\ \forall \x\in\mathcal{R}$. 
This is the typical situation of truncated samples.
A softer version is given by $\Omega(\x)<1$, which occurs when observations suffer from under-reporting.%
\footnote{Without a loss in generality, we will assume that over-reporting has been properly corrected, so that $\Omega(\x)\leq1$, e.g. by $\Omega(\x) \rightarrow \Omega(\x) \big / \max_{\x}\lbrace \Omega(\x) \rbrace$. We implicitly assume that there are not false positives in the data.}

Before we adapt the EM algorithm to incorporate $\Omega$, it is instructive to clarify how our concept of completeness relates to the terminology of ``missingness'' introduced by \citet{Rubin76.1} for missing features.
In \autoref{sec:missingness} we argue that the situation we find here is equivalent to ``missing at random'' (MAR), which allows for an arbitrary dependence of $\Omega$ on $\x$.
This includes $\Omega(\x) = \text{const.}$, a case called ``missing completely at random'' (MCAR), which is irrelevant for density estimation \red{because it is cancelled from \autoref{eq:observed_gmm}.}
Thus, MCAR data have the same expectation value of $p(\x)$ as completely observed data, the estimate is merely constrained by fewer samples.

On the other hand, the data are not allowed to conform to the most general case called ``missing not at random'' (MNAR), which would arise from a relation between the missing data and the density itself, $\Omega(\x)\rightarrow \Omega\left(\x, p(\x)\right)$ \citep[see e.g.][]{Schafer02.1}. A generalization, \red{which cannot use the factorization $p_o(\x)\propto \Omega(\x)\,p(\x)$ of \autoref{eq:observed_gmm},} goes beyond the scope of this work.

As the form of the likelihood function $\like$ under MAR is unchanged \red{up to a normalization constant}, the E-step remains unchanged as well, and all corrections appear in the M-step. 
\citet{Lee12.1} demonstrated that the effects of simple truncation of a GMM can be addressed by computing the zeroth, first, and second moments of each Gaussian component when truncated in the same way as the data, which requires the analytic integration of $p_k$ within the observed bounds. 
Essentially, the update equations for $\bmu_k$ and $\bSigma_k$ get a correction term from the difference between the current-iteration parameters and their truncated values.
As we seek the ability to employ arbitrary completeness functions, with complex spatial shapes and probabilistic rejection, the analytical integration becomes cumbersome, so we prefer approaches that draw samples from $\Omega$.

\subsubsection{Stochastic EM}

We adapt the Stochastic EM \citep{Diebolt1996} concept of augmenting missing data features by drawing imputing samples from the current model.
In our case, entire samples can be unobserved, as opposed to some of their features being missing. 
\red{
We thus need apply the selection process described by $\Omega$ and keep the rejected samples, i.e. we perform reverse rejection sampling with the current GMM as the proposal distribution.

In detail, we draw $S$ samples from the GMM and split them into those that we would have observed $\mathcal{O}$ and those that would be missing $\mathcal{M}$. The combined distribution has, by construction, the distribution $\Omega(\x)\,p(\x) + \left(1-\Omega(\x)\right)\,p(\x) = p(\x)$,
for which the previous EM equations (\autoref{eq:EM1} for the noise-free case, and \autoref{eq:EM2} for the noisy case) naturally hold.

If we adjust $S$, so that $|\mathcal{O}|$ is consistent with $N$ (the number of actually observed samples in $\mathcal{D}$), we can replace $\mathcal{O}$ with $\mathcal{D}$, and $S$ becomes the current estimate of the number of samples our data set would have had without rejection by $\Omega$. 
\autoref{eq:EM1} is extended thusly:

\begin{equation}
\label{eq:EM3}
\begin{split}
\text{E-step:}\ \ &q_{ik}\leftarrow \frac{\alpha_k p_k(\x_i)}{\sum_j \alpha_j p_j(\x_i)}\ \forall i \in \lbrace\mathcal{D}, \mathcal{M}\rbrace \\
\text{M-step:}\ \ &\alpha_k \leftarrow \frac{1}{N+|\mathcal{M}|} \Bigg(\sum_{i\in\mathcal{D}} q_{ik} +  \sum_{i\in\mathcal{M}} q_{ik}\Bigg) \equiv \frac{1}{N'} q_k\\
&\bmu_k \leftarrow \frac{1}{q_k} \Bigg(\sum_{i\in\mathcal{D}} q_{ik}\x_i +\sum_{i\in\mathcal{M}} q_{ik}\x_i\Bigg)\\
& \begin{split}
\bSigma_k \leftarrow \frac{1}{q_k}
\Bigg(&\sum_{i\in\mathcal{D}} q_{ik}\left[ (\bmu_k - \x_i)(\bmu_k - \x_i)^\top\right] +\\
&\sum_{i\in\mathcal{M}} q_{ik}\left[ (\bmu_k - \x_i)(\bmu_k - \x_i)^\top\right]\Bigg).
\end{split}
\end{split}
\end{equation}
Because of the linearity of the equations, the correction terms for the moments can be computed from $\mathcal{M}$ missing samples and added to the ones we compute for $\mathcal{D}$.
The normalization constant $Z$ can be obtained from the imputation sample by Monte Carlo integration,
\begin{equation}
\label{eq:obs-Z}
Z(\btheta,\Omega)\approx \frac{1}{S} \sum_{\x \in \lbrace\mathcal{D}, \mathcal{M}\rbrace} \Omega(\x),
\end{equation}
of which only the contribution from $\mathcal{M}$ has to be determined in each iteration.
In case of a binary $\Omega$, $Z\approx\tfrac{N}{S}$.}

This approach, which we will call \algo, in which at each iteration we draw and augment the observed data with imputation samples, is summarized in \autoref{alg:gmmis}. 
It is guaranteed to maximize the complete-data log-likelihood and converges to a stationary distribution for the parameters $(\balpha,\bmu,\bSigma)$ \citep{Diebolt1996,Nielsen00}.
To the best of our knowledge, this is the first time that a GMM approach has been made robust against not just truncation but the generalized form of incomplete sampling.

This method is efficient where $p(\x)\left(1-\Omega(\x)\right)$ is large; correction for weakly expressed components are more difficult to attain because $\mathcal{M}$ is drawn globally from the entire GMM, as opposed to being drawn from each component individually. 
A minor limitation is that when $\Omega$ is large for a component, the values of the $\mathcal{M}$-sums in the M-step equations are not well determined, however their impact on the result is also minor because that component is mostly observed.

In comparison with an analytic evaluation of the PDF in the unobserved regions, this approach is flexible and efficient, does not require numerical integrations of the component distributions, but, due to its stochastic nature, suffers from sample variance in the correction terms because $\mathcal{M}$ has finite size.
As with all Stochastic EM approaches, it generates sequences of $\mathcal{L}_o$ that are not guaranteed to monotonically increase \emph{in every step}.
To reduce the stochastic contribution to $\mathcal{L}_o$, we can average the correction terms from $\mathcal{M}$ over multiple draws of size $S$, so that the corrections become more precise and  $\mathcal{L}_o$-sequences closer to monotonic.
We found that averaging over $\approx 10$ imputation samples results in sample variances that are small compared to the increase of the observed likelihood.

\subsubsection{Incompleteness and Noise}
Besides speed and flexibility, \algo\ retains the ability to deal with noisy samples. By adding noise to the imputation samples $\mathcal{M}$, we can evaluate $q_{ik}$, $\bb_{ik}$, and $\bB_{ik}$ for $i\in\mathcal{M}$ and modify \autoref{eq:EM2} analogously to \autoref{eq:EM3}:

\begin{equation}
\label{eq:EM4}
\begin{split}
\text{E-step:}\ \ &q_{ik} \leftarrow \frac{\alpha_k p_k(\y_i \mid \bmu_k,\bT_{ik})}{\sum_j \alpha_j p_j(\y_i \mid \bmu_j,\bT_{ij})} \ \forall i \in \lbrace\mathcal{D}, \mathcal{M}\rbrace\\
&\bb_{ik} \leftarrow \bmu_k + \bSigma_k\bT_{ik}^{-1}(\y_i-\bmu_k) \ \forall i \in \lbrace\mathcal{D}, \mathcal{M}\rbrace\\
&\bB_{ik} \leftarrow \bSigma_k - \bSigma_k\bT_{ik}^{-1}\bSigma_k \ \forall i \in \lbrace\mathcal{D}, \mathcal{M}\rbrace \\
\text{M-step:}\ \ &\alpha_k \leftarrow \frac{1}{N+|\mathcal{M}|} \Bigg(\sum_{i\in\mathcal{D}} q_{ik} +  \sum_{i\in\mathcal{M}} q_{ik}\Bigg) \equiv \frac{1}{N'} q_k\\
 &\bmu_k \leftarrow \frac{1}{q_k} \Bigg(\sum_{i\in\mathcal{D}} q_{ik}\bb_{ik} + \sum_{i\in\mathcal{M}} q_{ik}\bb_{ik}\Bigg)\\
& \begin{split}
\bSigma_k \leftarrow \frac{1}{q_k}
\Bigg(&\sum_{i\in\mathcal{D}} q_{ik}\left[ (\bmu_k - \bb_{ik})(\bmu_k - \bb_{ik})^\top\right] +\\
&\sum_{i\in\mathcal{M}} q_{ik}\left[ (\bmu_k - \bb_{ik})(\bmu_k - \bb_{ik})^\top\right]\Bigg).
\end{split}
\end{split}
\end{equation}
As before, the corrections from $\mathcal{M}$ have the same form as, and are added to, the moments calculated for $\mathcal{D}$.
But to do so, we have to define $\bT_{ik}=\bSigma_k+\bS_i$ for the missing samples, which which we lack \emph{observed} uncertainties. These uncertainties may be known even in unobserved regions. If not, we need to make a guess of $\bS(\x)$ in the unobserved region, e.g. from the mean or a smooth interpolator of the observed $\bS_i$.

We must stress that not assuming errors for $\mathcal{M}$ would result in it having too large a weight in the likelihood. In the E-step, samples from $\mathcal{M}$ would be evaluated without noise, while the observed samples have a broadened likelihood under noise.
As a result, the EM algorithm would become dominated by missing samples and yield density estimates that are unduly shifted towards regions of low $\Omega$.
On the other hand, we do not have to exactly match the uncertainties of $\mathcal{M}$ to those of $\mathcal{D}$ because they only enter as sums in the M-step.
Matching the average errors of the samples associated with each component, instead of each individual sample, is therefore sufficient to estimate the parameters of all components. 

We finally note that adding noise and applying $\Omega$ do not commute; in either order, \algo\ requires only that $\mathcal{M}$ can be created such that it completes the data with an estimate of the unobserved portion. 

\begin{algorithm}[t]
\caption{\algo\newline
{\small A GMM is fit to observed samples $\mathcal{D}=\lbrace\x_i\rbrace_{i=1,\dots,N}$, accounting for a specified sample completeness $\Omega$ by drawing imputation samples $\mathcal{M}$ from the current-iteration GMM and accepting those with a rate $1-\Omega$. 
For noisy observations with covariances $\lbrace\bS_i\rbrace$, \algo\ requires an error model $\bS(\x)$ for the entire data region.}}
\label{alg:gmmis}
\begin{algorithmic}[1]
\Procedure{\algo}{$\lbrace\x_i\rbrace, \Omega(\x), \mathrm{tol},\, \left[\lbrace\bS_i\rbrace, \bS(\x)\right]$}
\For{$t  = 1,2,\dots$}
\State $\mathcal{Z}^{t } \gets \lbrace \z_i \sim p(\x \mid \balpha^t ,\bmu^t ,\bSigma^t ) \rbrace_{i=1,\dots,S^t}$
\State $\mathcal{R}^t \ \ \gets \lbrace r_i\sim\mathcal{U}(0,1)\rbrace_{i=1,\dots,S^t}$
\State $\mathcal{M}^t  \gets \lbrace \z_i : r_i < 1-\Omega(\z_i)\rbrace_{i=1,\dots,S^t}$
\If{$S^t - |\mathcal{M}^t| \nsim \mathrm{Poisson}(N)$}
  \State $S^t \gets S^t (S^t - |\mathcal{M}^t|)/N$
  \State go to Line 3 
\EndIf 
\If{$\lbrace\x_i\rbrace$ noise-free}
  \State $q^{t +1} \gets$ \ \autoref{eq:EM3} (E-step)
  \State $\balpha^{t +1}, \bmu^{t +1}, \bSigma^{t +1} \gets$  \ \autoref{eq:EM3} (M-step)
\Else
  \State $\bS_\z^t  \gets \lbrace \bS(\z_i)\rbrace_{\z_i\in\mathcal{M}^{it}}$
  \State $\mathcal{M}^t  \gets \lbrace \z_i^\prime \sim \mathcal{N}(\z_i, \bS^t _{\z_i})\rbrace_{\z_i\in\mathcal{M}^{it}}$
  \State $q^{t +1}, \bb^{t +1}, \bB^{t +1} \gets$ \ \autoref{eq:EM4} (E-step)
  \State $\balpha^{t +1}, \bmu^{t +1}, \bSigma^{t +1} \gets$ \  \autoref{eq:EM4} (M-step)
\EndIf
\State $\loglike_o^{t +1} \gets$ \autoref{eq:obs-likelihood} \& \autoref{eq:obs-Z}
\If {$|\loglike_o^{t +1} - \loglike_o^t|  < \mathrm{tol}\cdot \loglike_o^t$}
 break
\EndIf
\EndFor
\EndProcedure
\end{algorithmic}
\end{algorithm}

\section{Practical considerations}
\label{sec:practice}

\subsection{Initialization}
\label{sec:init}

The EM algorithm only guarantees convergence to a local maximum of the likelihood.
With the large number of free parameters of the GMM ($1+d+d(d+1)/2$ for each of $K$ components), a suitable initialization is critical.
Several initialization schemes have been proposed \citep[e.g.][]{Biernacki03.1,Bloemer13.1}.
For a completely observed data set, we adopt the simplest strategy, namely drawing the means at random from the data.
In detail, given a user-defined length scale $s$, for each component $k$ we draw $i$ at random from $\lbrace1,\dots,N\rbrace$ and $\Delta\x_i\sim \mathcal{N}(\vec{0}, s^2\tens{I})$, and set
\begin{equation}
\label{eq:init}
\begin{split}
&\alpha_k = 1/K\\
&\bmu_k = \x_i + \Delta\x_i\\
&\bSigma_k = s^2\tens{I},
\end{split}
\end{equation}
which naturally follows the distribution of the data on scales larger than $s$. 
To prevent strong initial localization, $s$ should be chosen to exceed the typical clustering scale of the data, but small enough that multiple components do not strongly overlap.

In the case of incomplete samples, we initially make the assumption that it was complete, i.e. $\Omega=1$, fit a GMM to the observed distribution, and use that fit to initialize the run with a specified $\Omega\neq1$.
While this approach will obviously fail if a component is entirely located in a region with $\Omega=0$, we found that an initial guess based on the observed distribution much more quickly converges than the random initialization described above.
To aid the exploration of the regions with $\Omega<1$, we leave the components means unchanged but multiply the covariances by a factor $> 1$.
If that factor is chosen too small, the EM algorithm will not be able to pick up the correction terms $\sum_{i\in\mathcal{M}}q_{ik}$ etc. in \autoref{eq:EM3} before re-converging to the previous, observed-sample location.
In turn, if the factor is chosen too large, the convergence is slowed down.
We therefore recommend a factor of 2 -- 4, i.e. increasing the linear size of each component by 50 -- 100\%, as a compromise that works well in practice.

\subsection{Split-and-merge operations}
\label{sec:snm}

With a large number of free parameters, the EM algorithm can easily get trapped in local maxima of $\like$. 
For clustered data, this behavior leads to GMM components being placed across several clusters or a single cluster being shared by multiple components. In the latter case, the weight $\alpha_k$ tends to zero for at least one of those components.

To improve the performance of the GMM, \citet{Ueda00.1} devised criteria to decide whether a component should be merged with another or be split into two.
Performing both of these operations at the same time amounts to altering three distinct components with the total number of components being conserved.
We follow this approach, with two alterations we found to perform better for all cases with $d=2,3$ we investigated.

\citet{Ueda00.1} proposed to merge the components $k$ and $l$ that maximize
\begin{equation}
J_\mathrm{merge}(k,l) = Q_k^\top\,Q_l\mathrm{,}
\end{equation}
with $Q_k^\top = (q_{1k},\dots,q_{Nk})$, i.e. the components whose posterior cluster assignment $p(k \mid \x)$ are most similar across the entire data set.
This works well in practice as long as there are no ``empty'' components, which do unfortunately arise in multimodal situations if more components are locally available than are needed to explain the data.
Because $q_{ik}\rightarrow 0$ when $\alpha_k\rightarrow 0$, the merge criterion above will not seek to merge such empty components even though they are obviously excellent choices to be merged.
We therefore replace $Q_k\rightarrow Q_k\big/\alpha_k$, which means that we seek to merge components whose $p_k(\x)$ is most similar for the entire data set.

For the split criterion, we found that the one suggested by \citet{Ueda00.1}, based on the Kullback-Leibler divergence, is unduly affected by outliers and often leads to split candidates that do not improve the likelihood.
Instead, we took guidance from a proposal of \citet{Zhang03.1} on how to best re-initialize the two new components that result from a split, namely to separate their means along the semi-major axis of the ellipsoid described by the pre-split $\bSigma$.
With this intuition it is natural to identify split candidates according to their largest eigenvalue $\lambda_{1}$ of $\bSigma$.
When searching for the component $k=\mathrm{argmax}_k\lbrace\lambda_{k,1}\rbrace$,  we assume it is strongly elongated because it seeks to describe two clusters at once.
The main failure mode of that split criterion is again related to (almost) empty components.
Their parameters, in particular $\bSigma_k$, are only poorly determined. Some of them are erroneously large and would thus be identified as split candidates, while they constitute much better merge candidates.
We thus propose to identify split candidates by selecting $k$ as the one that maximizes
\begin{equation}
J_\mathrm{split}(k) = \alpha_k \lambda_{k,1}.
\end{equation}
While the original split criterion of \citet{Ueda00.1} will seek to eliminate \emph{any} deviation of the local density from its approximation by a Gaussian-shaped component and is therefore generally applicable, our criterion appears to perform better for identifying a prominent failure mode for unconstrained GMMs: components that merge two clusters.
As the split is only accepted if it leads to an overall increase in the likelihood, it does not lead to increased fragmentation from penalizing the largest components.
Following \citet{Zhang03.1}, we then replace the means of two newly split components $l$ and $m$ as $\bmu_{l,m}=\bmu_k \pm \tfrac{1}{2} EV_1(\bSigma_k)$, i.e. along the primary eigenvector of the covariance matrix.

When dealing with incomplete data, we have not found split-and-merge operations to be problematic.
They are most useful when enabled during the initialization run as described in \autoref{sec:init}, which will then not be plagued by strong failure modes of the EM algorithm.
As a result, the imputation samples $\mathcal{M}$ will be more reliable, and the full algorithm will converge faster than without split-and-merge operations.

\subsection{Minimum covariance regularization}
\label{sec:w}

One problem of the objective function $\like$ is that it becomes unbounded if $\bmu_k = \y_i$ for any $k$ and $i$ because $\bSigma_k\rightarrow\mathbf{0}$. \citet{Bovy11.1} presented a regularization scheme to set a lower bound for every  component volume. In its simplest version it assumes the form of a $d$-dimensional sphere with variance $w$, which modifies the last update equation in \autoref{eq:EM2} according to
\begin{equation}
\label{eq:regularization}
\bSigma_k \leftarrow \frac{1}{q_k+1} \left[\sum_i q_{ik}\left[ (\bmu_k - \bb_{ik})(\bmu_k - \bb_{ik})^\top + \bB_{ik}\right] +w\tens{I}\right].
\end{equation} 
In principle, the value of $w$ is entirely arbitrary, but we choose to provide a more intuitive setting and associate it with a minimum scale $\omega$ in feature space, below which the model does not possess explanatory power, i.e. we want $\det(\Sigma_k) \geq \omega^{2d}$.
For a constant regularization term, we need to adopt a typical value of $q_k$, for which we use its mean over all components, $\bar{q_k}=\tfrac{N}{K}$.
We can thus set $w = \omega^2 (\tfrac{N}{K} + 1)$.
This choice does not provide an exact lower bound for each component as we determine the regularization term from the average value of $q_k$, which leads to stronger regularization for components with small $\alpha_k$.
We consider this an advantage in noisy situations, where the parameters of weakly expressed components can be hard to determine.

We note, however, that the form of the regularization in \autoref{eq:regularization} prefers features in the data to be of approximately equal size, otherwise the penalty term can dominate for small features while being ineffective for large features.

\subsection{Fitting for a background distribution}
\label{sec:background}

In many situations, observed data comprise anomalous samples that appear unclustered, i.e. unrelated to the features of interest, and rather originate from a more uniform ``background'' distribution.
One solution within the context of GMMs would be to add another component with very large variance and to fit for its amplitude only.
We prefer to introduce a specific background distribution over the relevant region $\mathcal{R}$ of feature space, 
e.g. the most conventional form of a uniform background
\begin{equation}
p_\mathrm{bg}(\x) = \begin{cases} 
\ \left[\int_\mathcal{R}d\x\right]^{-1}=\text{const.} & \mathrm{\x\in\mathcal{R}} \\
\ 0 & \mathrm{\x\notin\mathcal{R}.}
\end{cases}
\end{equation}
If the amplitude of the background component $\nu$ is unknown, one can introduce a two-level mixture model for the combined density distribution \citep[e.g.][their section 7.2.4]{Fruehwirth06}, 
\begin{equation}
p(\x \mid \balpha,\bmu,\bSigma, \nu) = (1-\nu) \sum_{k=1}^K \alpha_k\, p_k(\x) + \nu p_\mathrm{bg}(\x),
\end{equation}
but we prefer keeping the model strictly linear in the amplitudes and put the component and background amplitudes on equal footing:
\begin{equation}
p(\x \mid \balpha,\bmu,\bSigma, \nu) = \sum_{k=1}^K \alpha_k\, p_k(\x) + \nu p_\mathrm{bg}(\x).
\end{equation}
In this form, $\sum_k \alpha_k=1-\nu$.
To determine the amplitude $\nu$ we require another indicator variable $q_{i\text{bg}}$ to denote if a sample $\x_i$ belongs to the background.
Analogous to $q_{ik}$ in \autoref{eq:EM1} it is given by the posterior of $\x_i$ under the background model, which leads to these E-step equations
\begin{equation}
\begin{split}
q_{ik} &\leftarrow \frac{\alpha_k p_k(\x_i)}{\sum_k \alpha_k p_k(\x_i) + \nu p_\mathrm{bg}(\x_i)}\\
q_{i\text{bg}} &\leftarrow \frac{\nu p_\mathrm{bg}(\x_i)}{\sum_k \alpha_k p_k(\x_i) + \nu p_\mathrm{bg}(\x_i)}.
\end{split}
\end{equation}
The M-step for the background amplitude is
\begin{equation}
\nu\leftarrow \frac{1}{N}\sum_i q_{i\text{bg}},
\end{equation}
while the M-step of the GMM components remains unchanged.

If the samples are noisy, the previous equations in this section hold, but we need to marginalize over the positions of the unobserved noise-free samples \citep{Bovy11.1} and modify the background distribution as
\begin{equation}
p_\text{bg}(\y_i \mid \bS_i) = \int d\x\ p_\text{bg}(\x) \mathcal{N}(\y_i \mid \mathbf{x},\bS_i), 
\end{equation}
which is equivalent to the change to the GMM $q_{ik}$ in \autoref{eq:EM2} compared to the noise-free case in \autoref{eq:EM1}.
For the uniform background distribution the marginalization amounts to the zeroth moment of the truncated multivariate normal distribution \citep[e.g.][]{Manjunath2012}.

One could think that sample incompleteness does not affect the inference of the background component because any information how $\Omega(\x)$ acts on samples drawn from $p_\mathrm{bg}$ is already entirely contained in $\Omega(\x)$ itself.%
\footnote{The presence of a non-vanishing background intensity can therefore in principle be used to estimate $\Omega(\x)$. However, for this work we require $\Omega$ to be known a priori.}
The problem with this notion is that we do not know the relative amplitudes of signal and background in the unobserved regions, which will vary because the signal does. 
For consistent results, we therefore create the imputation sample $\mathcal{M}$ by drawing from the GMM \emph{and} the background model according to the current-iteration value of $\nu$, and proceed as in \autoref{sec:gmmis}.

We caution that the introduction of a background component leads to additional uncertainties and a higher-di\-men\-sional parameter space. In particular, during the first iterations of the EM algorithm, the parameters of the GMM often only provide rather poor description of the data, so that many samples will have higher probability under the background model, resulting in few samples left to fit for the GMM parameters in the next iteration.
This failure mode highlights the importance of a suitable initialization when working with a background model, especially when its intensity becomes dominant.
To this end, we found in our tests that the $k$-means initialization from \citet[their Algorithm 1]{Bloemer13.1} performed more robustly than the random initialization of \autoref{eq:init}.

\subsection{Averaging estimators}
\label{sec:stacking}

Even with well-chosen initial values and split-and-merge operations, any single GMM will get trapped in local maxima of the likelihood.
We therefore advocate, for two reasons, to average several GMMs fit to the same data, a technique also known as ensemble learning.

First, ensemble estimators typically outperform even the best single estimator.
In particular, \citet{Smyth99.1} built an improved estimator by employing the ``stacking'' method proposed by \citet{Wolpert92.1}, which uses the cross-validation result to determine non-negative weights for each model \citep[see also][]{Breiman96.1}.
In the context of GMMs, it is useful to realize that a mixture of mixture models is still a mixture model, therefore no conceptual changes have to be made when evaluating the estimator.

Second, stacked GMMs reduce the need to determine the optimal number of components $K$.
The main concern when determining $K$ is that if set too high, the resulting GMM overfits the data, following spurious features that increase the variance in the prediction, while if set too low, the model does not capture essential features of the data, leading to a large prediction bias.
As this topic has been extensively discussed elsewhere, we will not address it here.
We do note, however, that stacking exhibits better trade-offs between bias and variance than single-model selection or uniform averaging, rendering stacked GMM particularly robust against overfitting \citep{Smyth99.1}.
Even more so, combining GMMs with different $K$, or different covariance constraints, can provide a natural framework to describe data with a variety of spatial scales.

A decision if and how single-run GMMs are to be averaged will depend on the characteristics of the data at hand.
In this work, we will only use single-run results to allow the reader to evaluate the performance of \algo\ rather than that of the averaging scheme.

\begin{figure*}[t!]
\includegraphics[width=0.33\linewidth]{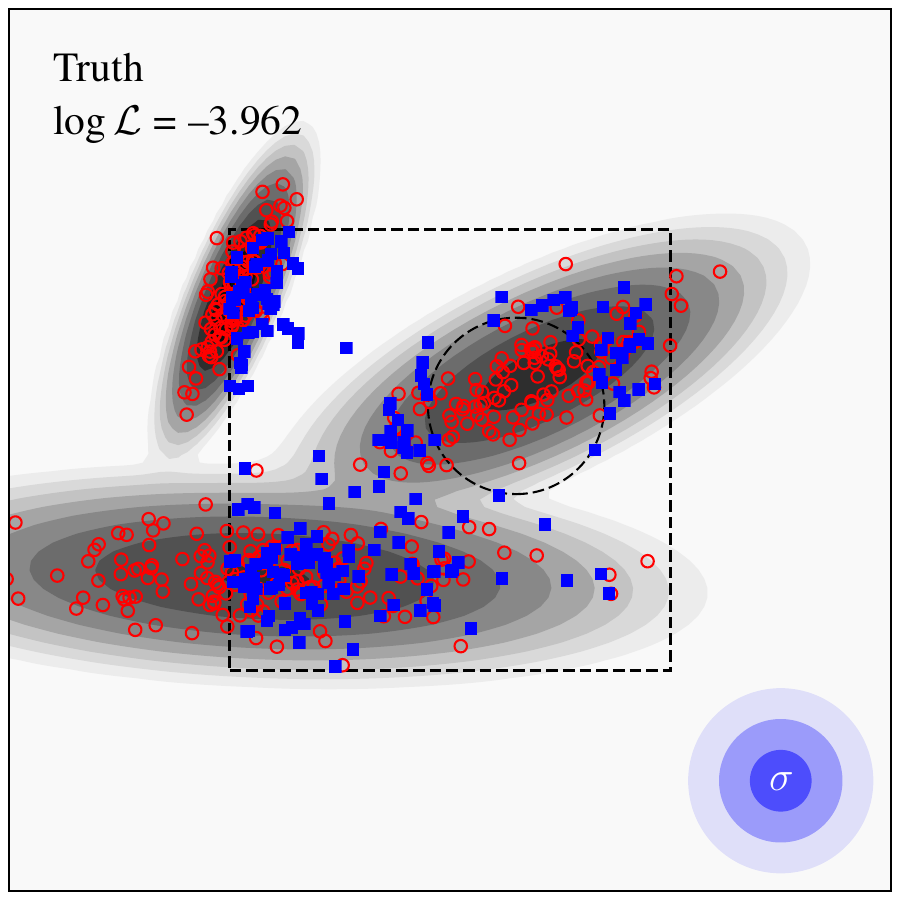}
\includegraphics[width=0.33\linewidth]{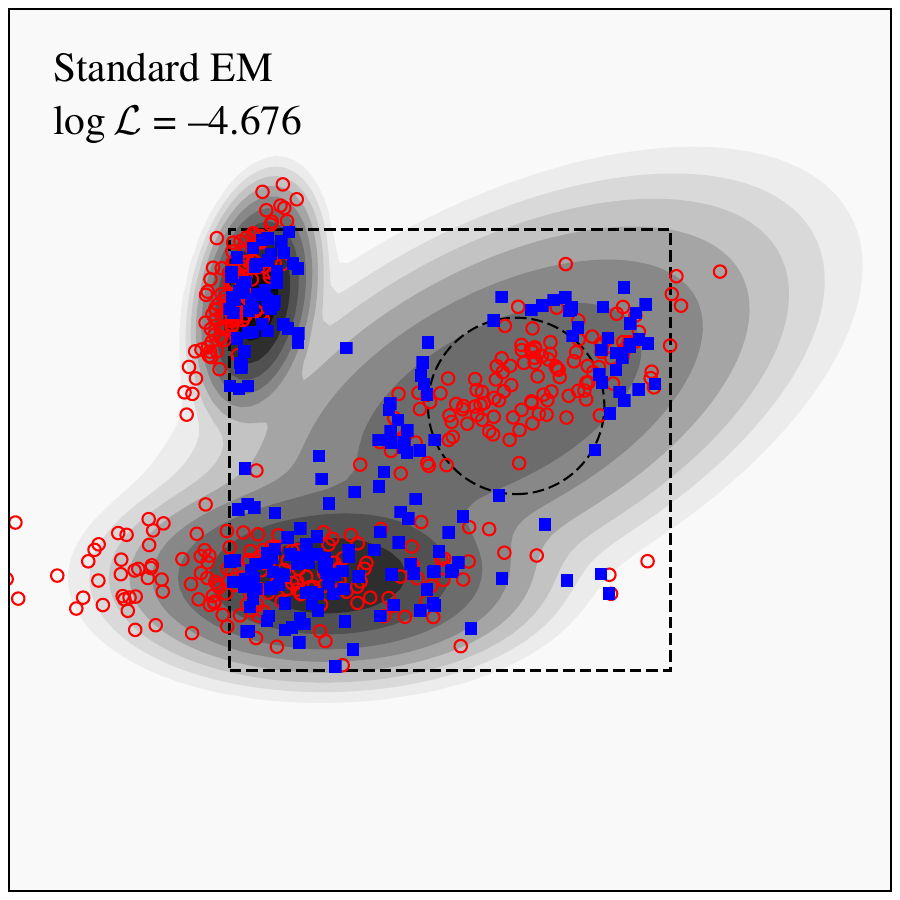}
\includegraphics[width=0.33\linewidth]{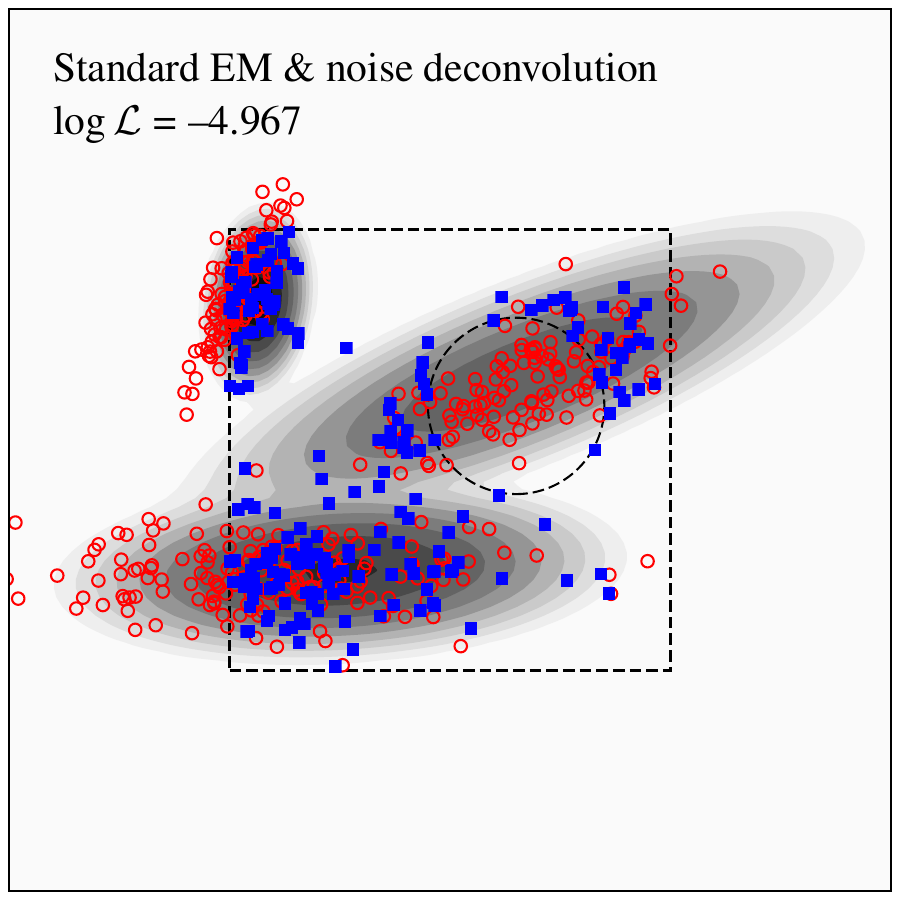}\\
\includegraphics[width=0.33\linewidth]{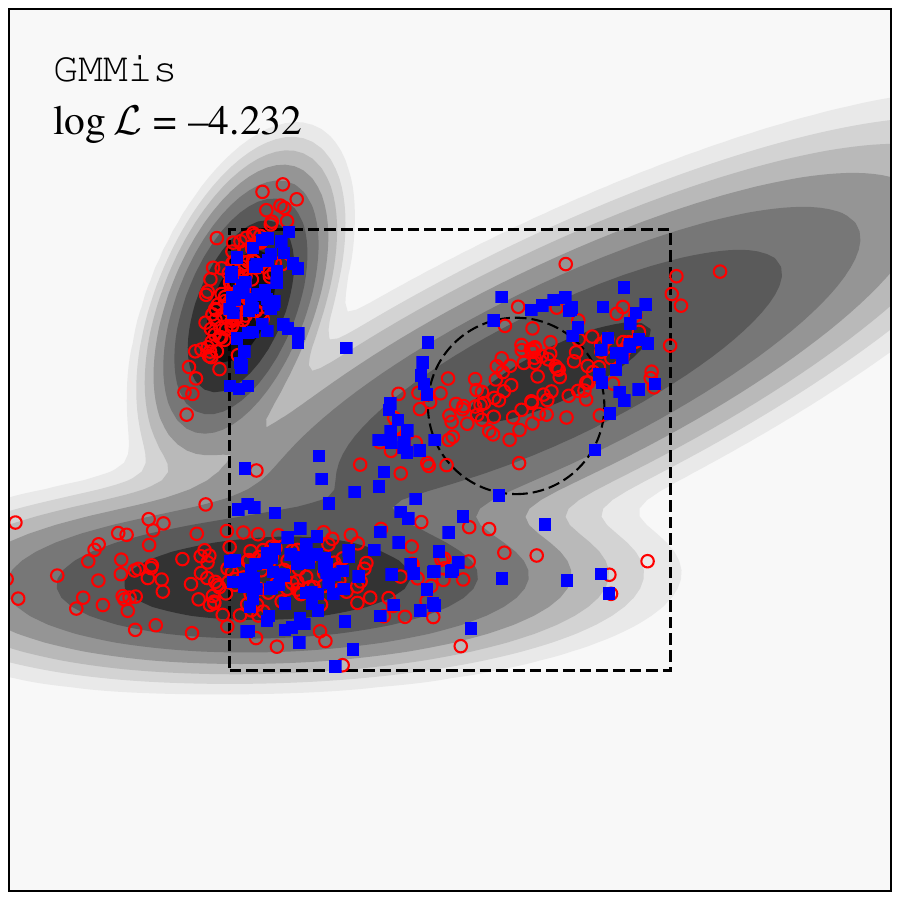}
\includegraphics[width=0.33\linewidth]{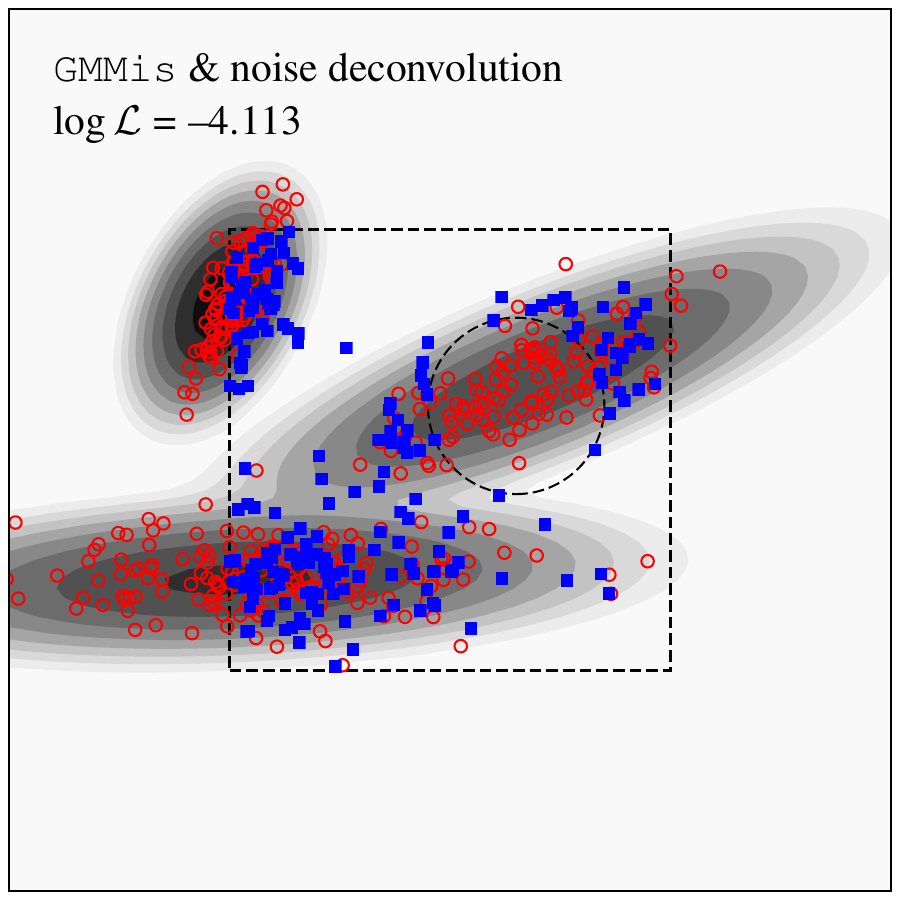}
\includegraphics[width=0.33\linewidth]{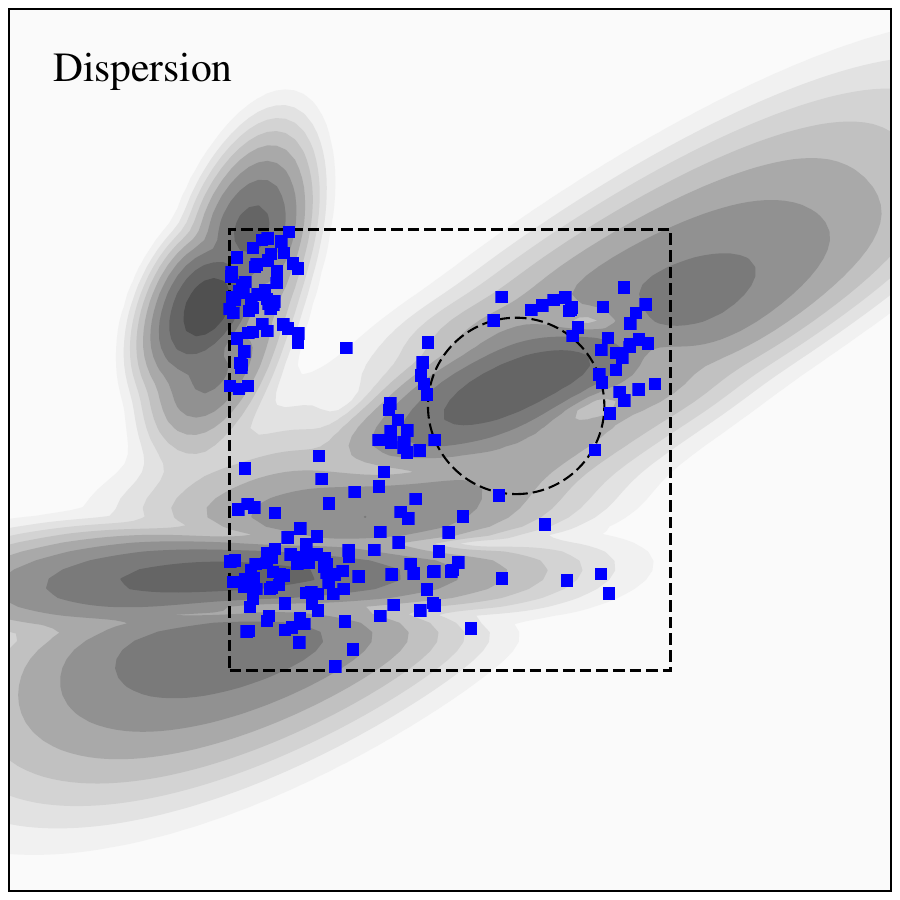}
\caption{Test case for the EM algorithm with noisy and incomplete samples. \emph{Top left}: True density distribution (contours in arcsinh stretch), a derived sample with $N=400$ (red open circles, whose average $\log\like$ under the respective models is given in the top-left corner of each panel), and the test sample (blue squares) after adding Gaussian noise (whose 1,2,3 $\sigma$ contours are shown in the bottom-right corner) and rejecting all points outside of the box or inside the circle (dashed curves).
\emph{Top center}: Result of the standard EM algorithm (\autoref{eq:EM1}). 
\emph{Top right}: Result of the standard EM algorithm with noise deconvolution (\autoref{eq:EM2}).
\emph{Bottom left}:  Result of the proposed EM algorithm \algo\ (\autoref{eq:EM3}), assuming noise-free samples. 
\emph{Bottom center}: Result of \algo, accounting the noise of the $\mathcal{O}$ and $\mathcal{M}$ samples. 
\emph{Bottom right}: Standard deviation of the predicted $p(\x)$ from 10 runs of \algo\ as in the bottom-center panel.
Markers, if present, and completeness boundaries are identical in all plots.
}
\label{fig:box}
\end{figure*}

\section{Experiments}
\label{sec:experiments}

\subsection{A toy example}

We first perform a simple test with no model misspecification, i.e. we draw the original sample from a GMM with $K=3$. In \autoref{fig:box}, we show that sample as red circles.
We then add a noticeable amount of Gaussian noise (blue contours) and impose a purely geometric completeness $\Omega(\x)$, whose boundaries are given by a box and a circle (dotted curves), resulting in the test sample (blue squares).
With this test case we demonstrate how the EM algorithm reacts to noise and a non-trivial completeness.
The standard EM algorithm (top-center panel) does what one needed to expect, namely to describe the observed sample as is.
It clearly prefers the region inside of the boundaries and thereby misestimates amplitudes, locations, and covariances of the components affected by $\Omega$.

With the noise-deconvolution approach of \citet{Bovy11.1}, summarized in \autoref{eq:EM2}, one can attempt to recover the noise-free distribution. 
However, in this case (shown in the top-right panel of \autoref{fig:box} the resulting shrinkage of the components is detrimental to the overall likelihood because the model is now even more confined to the observed region, reducing its ability to also describe samples beyond that.
In essence, the model is biased as before but more confident in its correctness as the noise contribution has been removed.
We generically find it to be true that noticeable incompleteness will need to be addressed first, and only then can one properly deconvolve from the noise, as we demonstrate with the progressive improvement in the next two tests.

When the sample incompleteness from $\Omega$ is correctly specified and considered via \autoref{eq:EM3}, \algo\ recovers component centers, orientations, and amplitudes much better (bottom-left panel), however the covariances remain inflated as the presence of noise is not yet corrected for.
This level of model fidelity could have been achieved with analytical integration \citep{Wolynetz79.1,Lee12.1} or sample binning \citep{McLachlan88.1,Cadez02.1}, but here none of those operations were necessary. In addition, neither of the aforementioned approaches can account for incompleteness and noise.

In the bottom-center panel, we use the full algorithm, i.e. the noise deconvolution of \autoref{eq:EM2} applied to both the $\mathcal{O}$ and $\mathcal{M}$ samples in \autoref{eq:EM3}, resulting in the highest likelihood of the three variants, with just $\Delta\loglike=-0.151$ compared to red sample drawn from the true underlying PDF.

By running the full algorithm (with $\Omega$ and noise treatment) ten times, we can investigate the spread associated with the given data set under the admittedly restrictive assumption of a GMM with $K=3$. 
The bottom-right panel of \autoref{fig:box}, whose contours have the same stretch as those of the previous panels, shows that there are considerable differences between runs, mostly associated with the $\Omega$-boundaries.
Also, the components from different runs emphasize the importance of different apparent sub-clusters.
This behavior is caused by the added noise, which obscures the presence of small-scale features.
As we attempt to infer the noise-free PDF from noisy samples, the models will amplify small-scale density fluctuations.
Because of such spurious sub-clusters or, more generally, local minima of the likelihood, we advocate the use of more sophisticated averaging schemes, as outlined in \autoref{sec:stacking}.

\subsection{Limits of applicability}

\begin{figure}[ht]
\includegraphics[width=\linewidth]{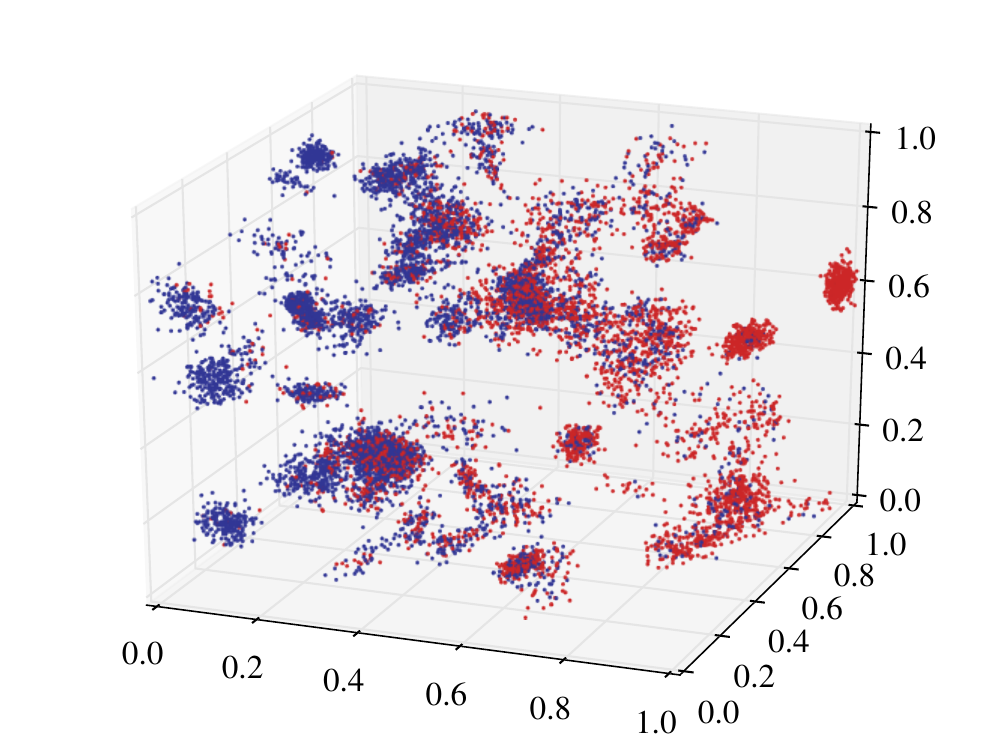}
\caption{Example test data generated from $K=50$ clusters in the $d=3$ unit cube. The completeness $\Omega$ is probabilistic, decreasing linearly from unity (left side of the cube) to zero (right side). Observed samples are shown in blue, missing samples in red.}
\label{fig:test_3d}
\end{figure}

\begin{figure}[ht]
\includegraphics[width=\linewidth]{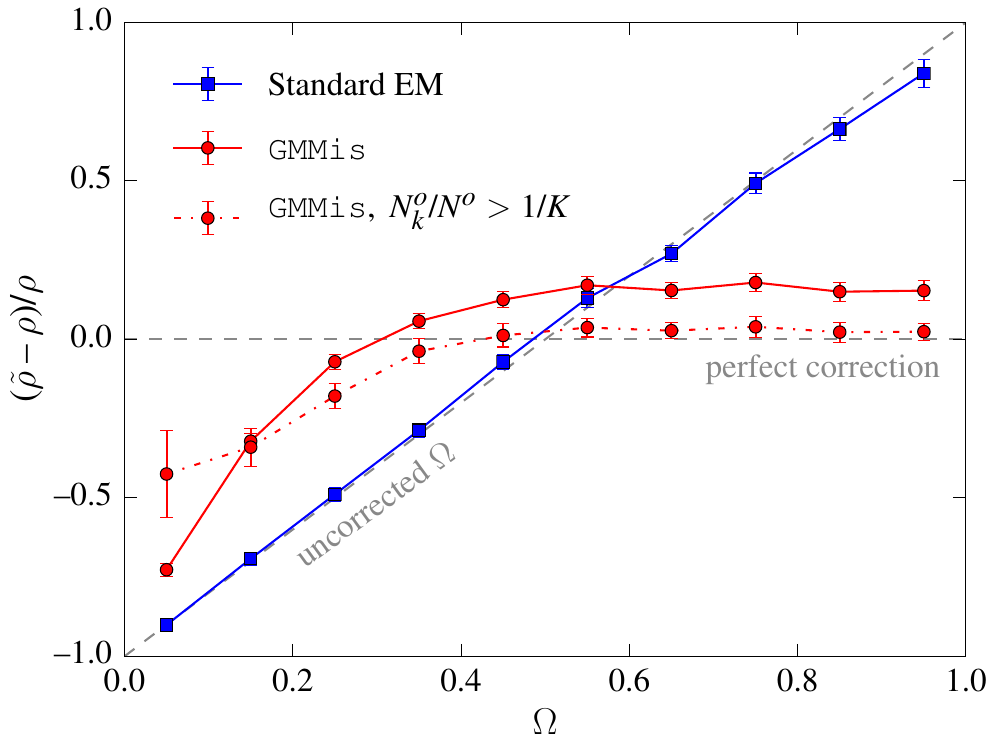}
\caption{Relative bias of the density estimator as function of completeness $\Omega$.
The density is calculated by generating $N=10,000$ samples from the fit and binning them into $50^3$ cells within the unit cube: for the standard EM algorithm (blue) which does not correct for incomplete samples; for \algo\ (solid red); for \algo\ after removing all poorly observed components from the input data that do not obey \autoref{eq:poor} (red dash-dotted).
}
\label{fig:test_3d_rho}
\end{figure}

\begin{figure}[ht]
\includegraphics[width=\linewidth]{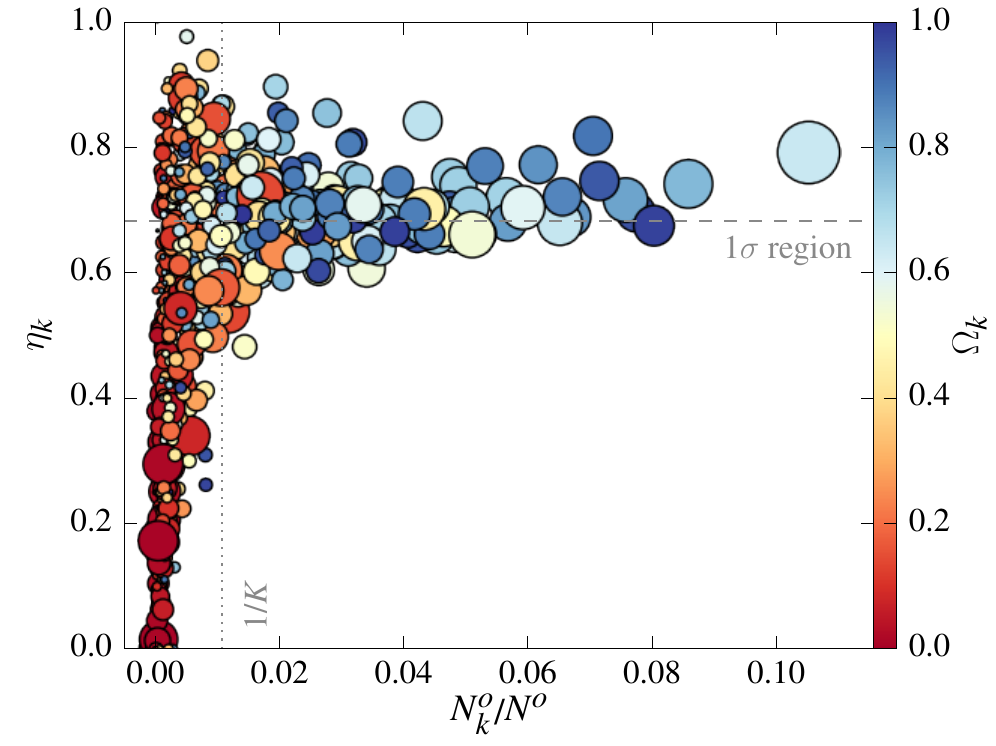}
\caption{Associated fraction $\eta_k$ of input samples to fit components as function of the effective weight in the observed data. Marker sizes denote weights $\alpha_k$ in the input data, colors refers to the average completeness $\Omega_k$ experienced by input components $k$. Both $N_k^\mathcal{O}$ and $\Omega_k$ are known exactly from the test setup. The horizontal dashed line shows the perfect outcome of 68.3\% of the samples found within the ``1$\sigma$ region'' of an output component; the vertical dashed line shows the criterion \autoref{eq:poor} for likely unrecoverable components.}
\label{fig:eta}
\end{figure}

We want to investigate, in a more difficult, highly multi-modal situation, how strongly incomplete the sampling can be for the proposed method to still yield reasonable density estimates.
We therefore set up a new test case in which we place $K=50$ GMM components randomly in the $d=3$ unit cube. Covariances are chosen such that the overlap between the components is not excessive; weights are drawn from a symmetric Dirichlet distribution with concentration parameter of unity. 
We impose a probabilistic completeness function $p(\Omega \mid \x) = 1 - x_1$, i.e. a linear ramp from fully observed to fully missing along one of the coordinate axes.
The purpose of this setup is to allow all combinations of component weight $\alpha_k$ and completeness $\Omega$.
We draw $N$=10,000 samples from the model, apply $\Omega$, and record the size of samples $N_k$ drawn from each component $k$ as well as the mean $\Omega_k$ experienced by those samples, so that we know the observed sample size $N^\mathcal{O}_k = N_k\Omega_k$. No noise is added to the samples. An example cube is shown in \autoref{fig:test_3d}.

We then fit the test data with another GMM with the same $K=50$, resulting in best-fit parameters $\tilde{\btheta}$, and repeat the process 10 times.
As diagnostic, we directly compare the densities $\rho(\x)$ by splitting the cube into $50^3$ cells and counting the input samples that fall into the cell covering $\x$.
We do the same for the predicted density $\tilde{\rho}(\x)$ by drawing $N$ samples from the fit.
As we know the expectation value of $\Omega$ at the location of each cell, we can compute the fractional bias in the predicted density  $\big(\tilde{\rho}(\x) -  \rho(\x)\big)\big/\rho(\x)$ as a function of $\Omega$ (\autoref{fig:test_3d_rho}).
The standard EM algorithm provides a reliable estimate of the \emph{observed} density, in other words: the bias is equal to $\Omega$.
The reason for positive bias for $\Omega>\tfrac{1}{2}$ lies in the normalization of the density estimator: as we drew the same number of samples $N$ from the model as where given as input data, any underestimation of the density must be compensated elsewhere.

In the same test, \algo\ shows a noticeably reduced sensitivity to $\Omega$, but the resulting density estimate is still biased.
This test demonstrates that there must be conditions under which the proposed algorithm fails.
It is obvious that we cannot expect \algo, or any other density estimator, to infer the properties or even existence of a cluster that is entirely unobserved, i.e. $N^\mathcal{O}_k = 0$.
Even for less extreme cases, \autoref{fig:test_3d_rho} shows that \algo\ struggles with samples from regions with low $\Omega$, overcompensating where $\Omega$ is large.

To determine limits of applicability, we evaluate how well the input samples from each component are described by the fit.
We compute the association fraction $\eta_k$ of the input sample from component $k$ that falls within ``1 $\sigma$'' of any component of the fit, which, for $d>1$, is more precisely expressed as $\lbrace i: (\x_i - \tilde{\bmu}_k)^\top \tilde{\bSigma}_k (\x_i - \tilde{\bmu}_k) \leq \chi^2_d(0.683)\ \text{for any}\ k\rbrace$, where $\chi^2_d$ is the quantile function of the chi-squared distribution with $d$ degrees of freedom.
For a component of the test data whose samples are perfectly fit (i.e. fit by the component from which they are generated), the association fraction should on average be $\eta_k=68.3\%$.
If $\eta_k$ is higher, either the associated component of the fit is too extended  or the test samples are associated with multiple components.
If $\eta_k$ is lower, the fit has effectively ignored some samples from input component $k$.
In \autoref{fig:eta} we plot $\eta_k$ as a function of the effective weight $N^\mathcal{O}_k\big/N^\mathcal{O}$ of component $k$ given the observed data, while the size of the marker encodes the initial weight $\alpha_k=N_k\big/N$, and the color encodes $\Omega_k$.
It is apparent that components of the test data with large $\alpha_k$ or $\Omega_k$ are generally well-described by the fit. 
For $\Omega_k\approx 0$, the fit will miss a large fraction of the original sample irrespective of $\alpha_k$, but there are also a few well-observed but low-$\alpha$ components that are largely being ignored by the fit.
There is also an increased tendency of fitting multiple nearby input clusters with one larger component, which leads to artificially large $\eta_k$ at low $N^\mathcal{O}_k\big/N^\mathcal{O}$.
Without formal proof, we seek a criterion to identify components that are too poorly observed to yield a reasonable fit. 
By performing the test described above with different values of $N$, $K$, component volumes, and Dirichlet concentration parameter, we found 
\begin{equation}
\label{eq:poor}
\frac{N^\mathcal{O}_k}{N^\mathcal{O}} = \frac{N_k\Omega_k}{\sum_k N_k\Omega_k}\gtrapprox \frac{1}{K}
\end{equation}
to perform well (shown as vertical dotted line in \autoref{fig:eta}).
This qualitatively agrees with findings of e.g. \citet{Naim12.1} that the best GMM fits with an EM algorithm are achieved when all components have approximately equal weights, in particular when they overlap.
It also implies that \algo\ has to successfully fit the observed distribution first to be able to generate meaningful imputation samples $\mathcal{M}$ that account and correct for $\Omega$.
If the first step is flawed, the second one will be, too.  

This two-step picture is confirmed when we remove from the test data all components that would have violated \autoref{eq:poor}, in other words, the test data now only comprises clusters that remain salient despite $\Omega$.
In this case, the algorithm can detect all present clusters and correct for the sample incompleteness.
Accordingly, the \algo\ density estimates are now largely unbiased (dash-dotted red line in \autoref{fig:test_3d_rho}).
We emphasize that this shortcoming is not caused by an incorrect form of the likelihood, rather by the economy of the EM algorithm to converge to the nearest likelihood maximum, which favors the most prominent clusters in the observed sample.
If, on the other hand, the parameters of all but one component were fixed at their true values and there were only one component left to fit, this component could experience incompleteness in excess of \autoref{eq:poor} and still be fit with a fidelity commensurate to its number of samples $N_k$.

\section{Application to \chandra\ X-ray data}
\label{sec:application}

\begin{figure*}[t!]
\includegraphics[width=0.5\linewidth]{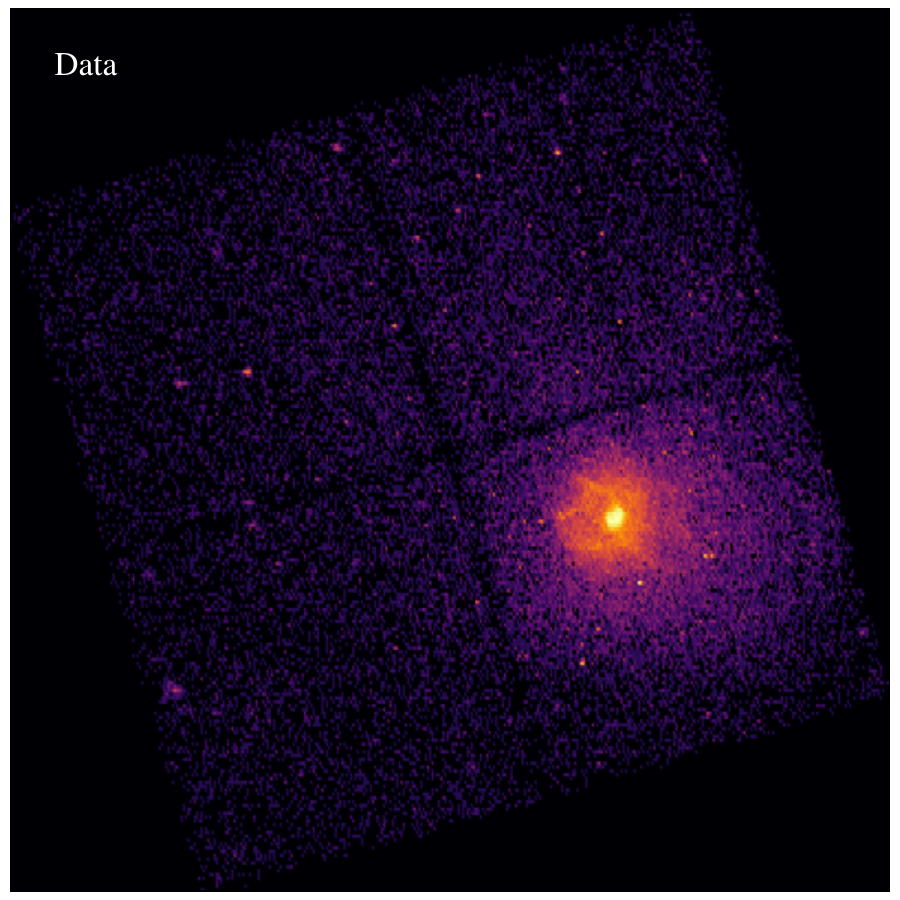}
\includegraphics[width=0.5\linewidth]{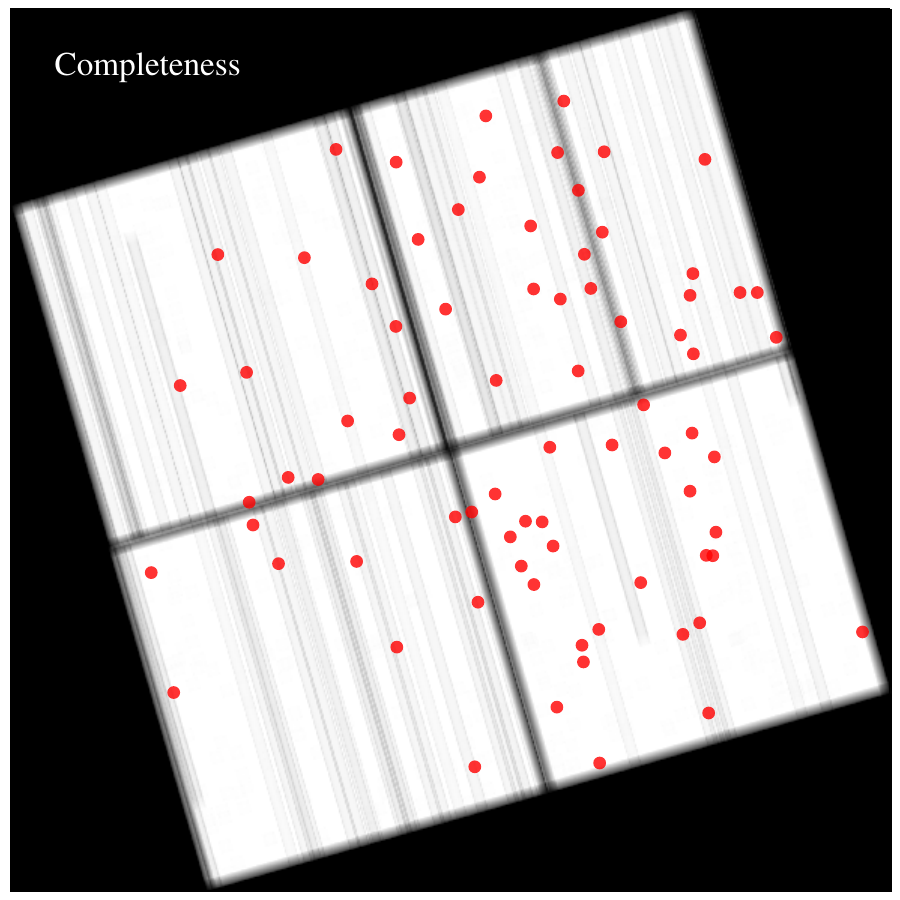}\\
\includegraphics[width=0.5\linewidth]{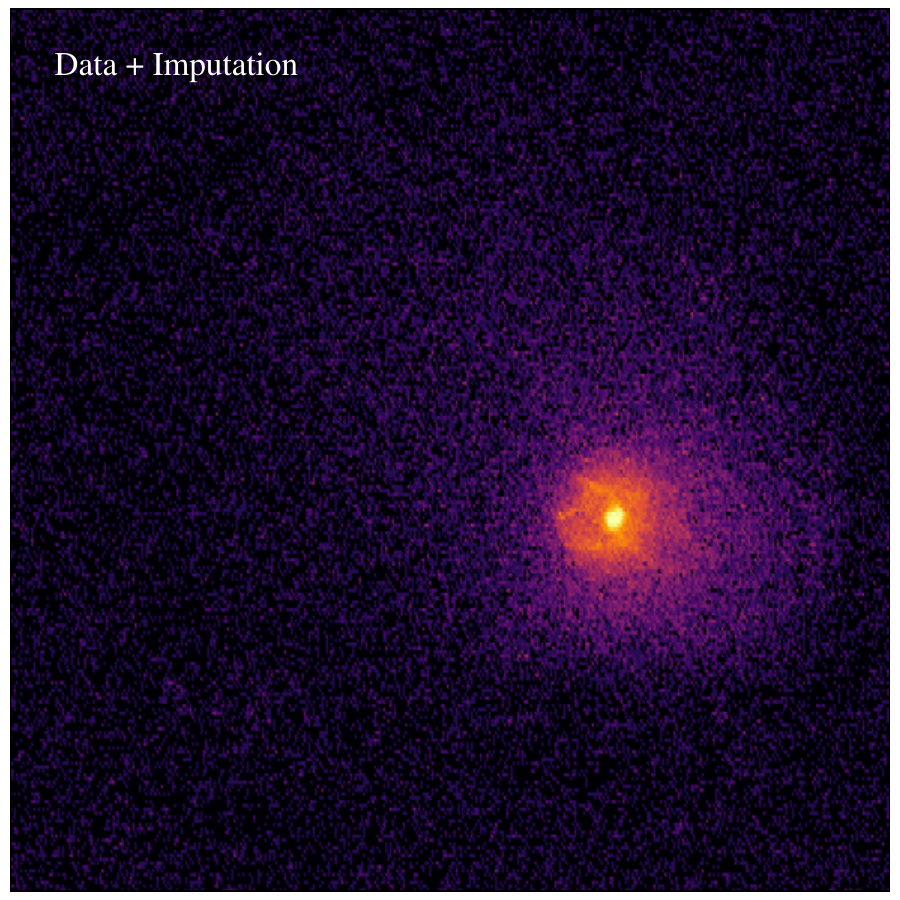}
\includegraphics[width=0.5\linewidth]{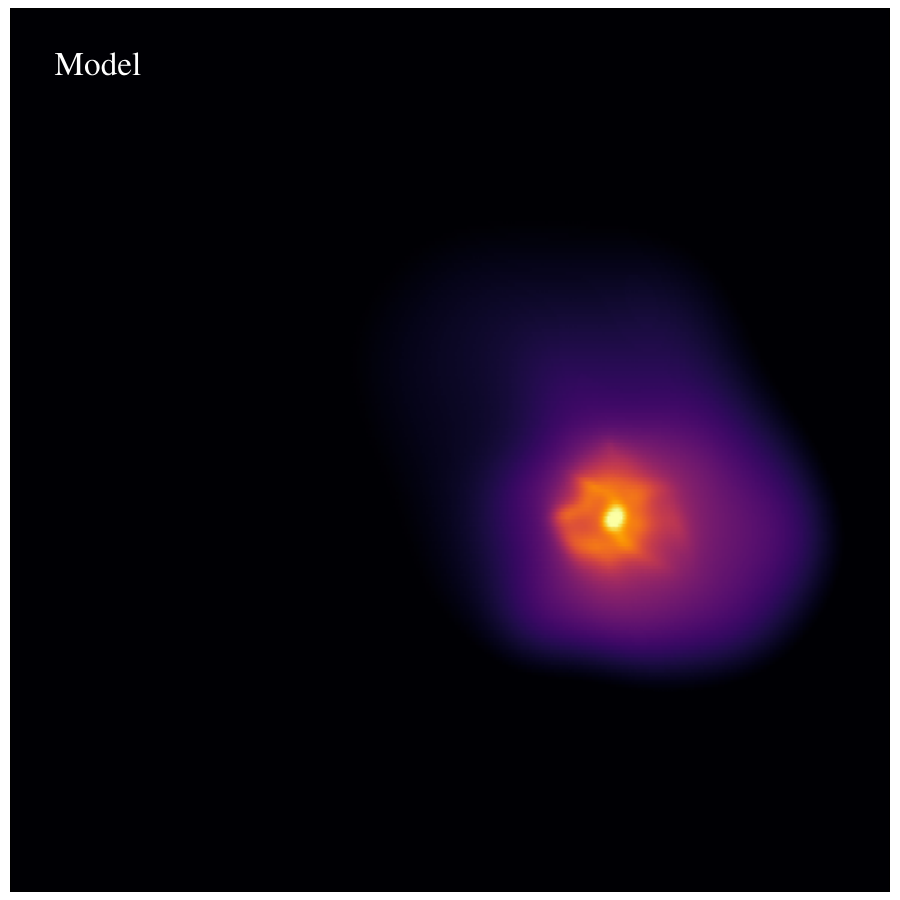}
\caption{Application to X-ray data. \emph{Top left}: Histogram of the locations of photons in the energy range 0.5--2 keV for an observation of the galaxy \object{NGC 4636} with the ACIS-I instrument of the \chandra\  X-ray telescope.
\emph{Top right}: Completeness $\Omega(\x)$ for ACIS-I, determined by the location of its four CCDs and detector sensitivity variations; point source masks (red circles, sizes true to scale), where we reject data and explicitly set $\Omega=0$, have been added for the purpose of this analysis. 
\emph{Bottom left}: ACIS-I data augmented with the imputation sample drawn from the final models of the GMM and the background.
\emph{Bottom right}: Final GMM with $K=60$ of the extended emission of the galaxy, after rejecting point sources, removing the background, and reconvolving to match the observation.
The images are aligned such that North is up and East is left; all panels are shown in logarithmic stretch.}
\label{fig:chandra}
\end{figure*}

To demonstrate the capabilities of \algo, we analyze the distribution of X-ray photons of the nearby galaxy \object{NGC 4636}, observed with the Advanced CCD Imaging Spectrometer (ACIS) aboard the \chandra\ telescope.
These data were retrieved from the \chandra\ public archive, and consist of two individual 75~ks pointings (Observation Identification Numbers 3926 and 4415). Basic data processing was carried out following the procedure described in \citet{Goulding16}. Briefly, we used the \chandra\ X-ray Center pipeline software packages available in {\sc ciao} v4.7 to apply the latest detector calibration files, and remove the standard pixel randomization, streak events, bad pixels, and cosmic rays. Photon catalogs (referred to as ``events files'') were screened using a typical grade set (grade = 0, 2, 3, 4, 6), and cleaned of $3 \sigma$ background flares. Finally, aspect histograms were constructed and convolved with the ACIS-I chip map, using the {\sc ciao} tool {\tt mkexpmap}, to generate the observation specific exposure time maps.

In the top-left panel of \autoref{fig:chandra} we show a histogram of $\approx$ 150,000 photons in the energy range $E\sim 0.5-2$~keV, covering an area with a side-length of about 0.3 degrees.
Two features of the observation are obvious. 
First, there are small gaps between the four CCDs of ACIS-I, where the ability to record photons is strongly reduced.
In detail, ACIS additionally suffers from minor and well-known sensitivity degradations in different parts of the CCD (top-right panel of \autoref{fig:chandra}).
Second, besides the galaxy, there is an essentially uniform particle background, as well as additional smaller objects (``point sources''). These point sources are a mixture of X-ray binary systems that are intrinsic to \object{NGC 4636} and distant, rapidly growing supermassive black holes unrelated to the target galaxy.
We mask the point sources with small circular apertures (shown as red dots), a step that is commonly done in the analysis of X-ray data.

As a further complication, the photon positions are not known exactly as they have been convolved with the instrument Point-spread function (PSF), which for this ACIS-I has a shape that is very well approximated by a circular Gaussian\footnote{See the \chandra\ Proposers' Observatory Guide, \url{http://cxc.harvard.edu/proposer/POG/}} with a width that varies from 0.4 arcsec in the inner region of ACIS-I to 15 arcsec at the perimeter.

Our new method is ideally suited to directly analyze the photon event files.
We can account for chip gaps, field edges, sensitivity variations, and point-source masks with the basic \algo\ algorithm from \autoref{sec:gmmis}.
By recognizing that convolution with the PSF is formally identical to additive measurement noise, we can employ the deconvolution method from \autoref{sec:EM-noisy} to build a generative model of the underlying, noise-free distribution of X-ray photons, while simultaneously fitting for the X-ray background as described in \autoref{sec:background}.
In contrast,  a more traditional image analysis typically entails smoothing, which does not correctly account for any form of sample incompleteness, and a single deconvolution step, which does not incorporate the spatial variation of the PSF width.

The bottom-left panel of \autoref{fig:chandra} demonstrates the principle of operations. By drawing samples from the current state of the model, which comprised the GMM and the uniform background, convolving them with the spatially varying PSF, and selecting them according to $\Omega$, we get a sample to augment the observed data. Combining this imputation sample and the point-source masked data, we get a representation of the internal state of the model, in other words: its estimate how the data would look like if $\Omega=1$. In each iteration these augmented data enter \autoref{eq:EM4} to determine the best-fit model of the extended emission of the galaxy without point sources (bottom-right panel).

We initialize the GMM with $K=60$ components with means distributed according to a bivariate Gaussian, whose width $s$ was determined by fitting the galaxy with $K=1$ first.
Because the scene exhibits features of different scales, we set the component covariances to $\bSigma_k=4^{-l}s^2 \ \mathbf{I}$ with $l={0,\dots,5}$, and 10 components at each level $l$.
This tailored initialization is necessary despite the GMM's ability to adjust to arbitrary configurations because the rate of convergence is very slow for the weakly expressed features we are particularly interested in.
The background amplitude $\nu$ was allowed to vary between 0.2 and 0.5, with $\nu=0.396$ being the best-fit value, in other words about 40\% of the photons over the unobscured region shown in \autoref{fig:chandra} originate from the background.

The GMM is capable of faithfully describing large and small-scale features in the data: a bright, apparently bimodal central region, most likely arising due to the presence of an accreting supermassive black hole; an extended halo of low-intensity emission from diffuse, $10^5-10^7$ Kelvin hot gas with a noticeable skewness towards the lower right corner; and the location and intensity of several shock fronts caused by buoyantly rising pockets of plasma that were likely inflated during previous outbursts of the central supermassive black hole.
However, because of the complexity of the model with $K=60$, it is not guaranteed that particular features, e.g. shock fronts, are fit by a single component. This association could be made more clear if the analysis operated on a three-dimensional feature space of photon positions and energies, a study we leave for the future.

Because this is a demonstration of the capabilities of the method, we have not performed e.g. cross-validation tests to determine the optimal $K$. The model is the result of a single run with a number of components that appeared visually reasonable to capture the key features of the data. Its $\chi^2$ per degree of freedom over an area where the galaxy dominates the emission is at least 1.34,\footnote{For the estimate of the degrees of freedom we assumed that no parameters are learned from the data, yielding an upper bound on the dof. If all parameters were linearly independent, which is not an appropriate assumption for GMMs, the dof would be reduced by 3\%.} indicating that a larger number of components may be necessary to capture all significant features.

\section{Summary and conclusions}
\label{sec:conclusions}

We describe a novel extension of the EM algorithm to perform density estimation with Gaussian mixture models in situations where the data exhibits a known incompleteness $\Omega$.
The type of incompleteness may be described by a sharp boundary, a case that is usually denoted as ``truncated data'',  or by an arbitrary probabilistic function $\Omega(\x)$, as long as the mechanism that causes the incompleteness is independent of the density (\emph{missing at random}).

The key difference of the method described here to the situation that is often---and somewhat ambiguously---called ``missing data'' is that the latter refers to data, for which some features of any data sample may be absent.
In contrast, we deal with entire samples being potentially absent, caused by the systematic limitation to observe the entire feature space or all of its relevant regions. 

Our solution is based on drawing imputation samples from the current state of the GMM, and to recompute the model parameters in the M-step using both observed and missing samples.
This technique is applicable to any generative model that is fit with an EM algorithm, as demonstrated with the signal--background model of \autoref{sec:background}.
Its advantage is the flexibility to efficiently adjust to any incompleteness $\Omega(\x)$ because one avoids the analytic integration of the predicted density over the incomplete regions. 
Instead, we draw test samples from the current model, retain those that would not have been observed under $\Omega$, and obtain non-zero contributions to the moments of those components that extend into incompletely observed regions.
If the model is suitable to describe the data-generating process, adding those moments to the ones obtained from the observed samples results in parameter estimates consistent with those for completely observed data.

We detail practical refinements of the algorithm, regarding its initialization, split-and-merge operations, and simultaneously fitting for a uniform background. 
We also recommend averaging GMMs from independent runs to smooth out the dependence of the EM algorithm on the starting position to estimate the uncertainty of the estimators. 

In simplified tests we demonstrate that the algorithm
\begin{enumerate}
\item recovers an estimate of the underlying density in presence of complex incompleteness functions,
\item correctly accounts for incompleteness and additive noise if the imputation samples describe the \emph{noisy} unobserved data,
\item can only partially correct for incompleteness if the too many of the original samples have not been observed.
\end{enumerate}
The last finding is central to the applicability of the method.
Constraining complex GMMs with finite amounts of samples becomes even more challenging with non-neglibible incompleteness. Correcting for the latter requires being able to perform the former reasonably well, otherwise the imputation samples do not describe the true unobserved samples.
In absence of an analytical estimate of the fidelity of GMMs from the EM algorithm, we provide a best-effort estimate for the limits of the algorithm in \autoref{eq:poor}.

We demonstrate the usefulness of the algorithm with example data from the NASA X-ray telescope \emph{Chandra}, where the incompleteness stems from gaps between the chips of the ACIS instrument.
We directly estimate the extended emission of the galaxy \object{NGC 4636} from the location of photon hits, while accounting for the window-frame configuration of the detector; a spatially varying point-spread function; and a uniform X-ray background.

The presented method provides an efficient and flexible tool to estimate the density of a process that is affected by moderate levels of incompleteness of the MAR type.
As such situations may arise in many areas of the physical and social sciences, we believe this contribution to be of general use and have therefore made our \python\ implementation available at  \url{\repo}.

\appendix
\section{Missingness for density estimation}
\label{sec:missingness}

To understand what form the likelihood assumes when some samples are not observed, we need to determine how observed and missing samples are related.
For this purpose \citet{Rubin76.1} introduced the missingness mechanism $R$ that determines which part of the complete data $\mathbf{Y}$ is observed ($\mathbf{Y}_o$) or missing ($\mathbf{Y}_m$).
The relation between $R$ and $\mathbf{Y}_o$ or $\mathbf{Y}_m$ gives rise to three distinct types of missingness: 
\begin{equation*}
p(R \mid \mathbf{Y}) = \begin{cases}
p(R) & \text{missing completely at random (MCAR)}\\
p(R \mid \mathbf{Y}_o) & \text{missing at random (MAR)}\\
p(R \mid \mathbf{Y}_o, \mathbf{Y}_m)&  \text{missing not at random (MNAR)}
\end{cases}
\end{equation*}
Under MCAR, the missingness mechanism does not depend on the data at all; in turn, the data do not reveal properties of $R$. The key distinction between MAR and MNAR is whether $R$ depends only on observed or also on missing features.

For the problem of density estimation with incomplete samples $\x$, it is not immediately obvious what $\mathbf{Y}_o$ and $\mathbf{Y}_m$ correspond to. 
We take guidance from the close relation between density estimation and function approximation by positive linear operators \citep[e.g.][]{Ciesielsky91.1}. 
We therefore take $\X$ to be the entire feature space $\mathbb{R}^d$ and $\Y=\mathbb{R}$ the co-domain of a scalar function $p: \X\rightarrow\Y$, namely the PDF $p(\x)$. 
The data then span $(\X, \Y) = \mathbb{R}^{d+1}$, and $R$ determines at what locations $\x$ a value $y\in\Y$ is recorded.

In \autoref{sec:gmmis}, we introduced the completeness function $\Omega(\cdot)$, which is equivalent to $p(R \mid \cdot)$. 
Only a spatially uniform completeness function fulfills the MCAR condition, but is irrelevant for density estimation as the resulting probability density function is simply a renormalized version of the true PDF.
If $\Omega$ only depends on $\x$, the MAR condition applies because  $\X$ is by construction completely observed (only values of $\Y$ may be missing).
Technically, MAR still holds if $\Omega$ depends on $\x$ and the density $p(\x^\prime)$ at some other observed position $\x^\prime\neq\x$, for instance when the observed samples ``shadow'' some others.
For the sake of brevity, we have not listed this case in \autoref{sec:gmmis}.
Only if samples are missing because of their own value of  $y=p(\x)$, or because of their relation to other missing values $p(\x^\prime)$, we have MNAR. 
This can happen e.g. when an experimental device is locally invalidated by saturation or if samples are only recorded if their local abundance exceeds a certain threshold.
An investigation under which conditions the proposed approach can account for MNAR cases is beyond the scope of this paper.

\bibliography{references.bib}

\begin{thebibliography}{}
\makeatletter
\relax
\def\mn@urlcharsother{\let\do\@makeother \do\$\do\&\do\#\do\^\do\_\do\%\do\~}
\def\mn@doi{\begingroup\mn@urlcharsother \@ifnextchar [ {\mn@doi@}
  {\mn@doi@[]}}
\def\mn@doi@[#1]#2{\def\@tempa{#1}\ifx\@tempa\@empty \href
  {http://dx.doi.org/#2} {doi:#2}\else \href {http://dx.doi.org/#2} {#1}\fi
  \endgroup}
\def\mn@eprint#1#2{\mn@eprint@#1:#2::\@nil}
\def\mn@eprint@arXiv#1{\href {http://arxiv.org/abs/#1} {{\tt arXiv:#1}}}
\def\mn@eprint@dblp#1{\href {http://dblp.uni-trier.de/rec/bibtex/#1.xml}
  {dblp:#1}}
\def\mn@eprint@#1:#2:#3:#4\@nil{\def\@tempa {#1}\def\@tempb {#2}\def\@tempc
  {#3}\ifx \@tempc \@empty \let \@tempc \@tempb \let \@tempb \@tempa \fi \ifx
  \@tempb \@empty \def\@tempb {arXiv}\fi \@ifundefined
  {mn@eprint@\@tempb}{\@tempb:\@tempc}{\expandafter \expandafter \csname
  mn@eprint@\@tempb\endcsname \expandafter{\@tempc}}}

\bibitem[\protect\citeauthoryear{Biernacki, Celeux  \& Govaert}{Biernacki
  et~al.}{2003}]{Biernacki03.1}
Biernacki C.,  Celeux G.,   Govaert G.,  2003, \mn@doi [Comput. Stat. Data
  Anal.] {10.1016/S0167-9473(02)00163-9}, 41, 561

\bibitem[\protect\citeauthoryear{{Bl{\"o}mer} \& {Bujna}}{{Bl{\"o}mer} \&
  {Bujna}}{2013}]{Bloemer13.1}
{Bl{\"o}mer} J.,  {Bujna} K.,  2013, preprint, \href
  {http://adsabs.harvard.edu/abs/2013arXiv1312.5946B} {} (\mn@eprint {arXiv}
  {1312.5946})

\bibitem[\protect\citeauthoryear{{Bovy}, {Hogg}  \& {Roweis}}{{Bovy}
  et~al.}{2011}]{Bovy11.1}
{Bovy} J.,  {Hogg} D.~W.,   {Roweis} S.~T.,  2011, \mn@doi [Annals of Applied
  Statistics] {10.1214/10-AOAS439}, \href
  {http://adsabs.harvard.edu/abs/2011AnApS...5.1657B} {5, 1657}

\bibitem[\protect\citeauthoryear{Breiman}{Breiman}{1996}]{Breiman96.1}
Breiman L.,  1996, \mn@doi [Machine Learning] {10.1007/BF00117832}, 24, 49

\bibitem[\protect\citeauthoryear{Cadez, Smyth, McLachlan  \& McLaren}{Cadez
  et~al.}{2002}]{Cadez02.1}
Cadez I.~V.,  Smyth P.,  McLachlan G.~J.,   McLaren C.~E.,  2002, \mn@doi
  [Mach. Learn.] {10.1023/A:1013679611503}, 47, 7

\bibitem[\protect\citeauthoryear{{Ciesielsky}}{{Ciesielsky}}{1991}]{Ciesielsky91.1}
{Ciesielsky} Z.,  1991, Probability And Mathematic Statistics, \href
  {http://www.math.uni.wroc.pl/~pms/publicationsArticle.php?nr=12.1&nrA=1&ppB=%201&ppE=%2024}
  {12, 1}

\bibitem[\protect\citeauthoryear{{Dempster}, {Laird}  \& {Rubin}}{{Dempster}
  et~al.}{1977}]{Dempster77.1}
{Dempster} A.~P.,  {Laird} N.~M.,   {Rubin} D.~B.,  1977, Journal of the Royal
  Statistical Society. Series B (Methodological), \href
  {http://www.jstor.org/stable/2984875} {39, 1}

\bibitem[\protect\citeauthoryear{Diebolt \& Ip}{Diebolt \&
  Ip}{1996}]{Diebolt1996}
Diebolt J.,  Ip E.,  1996, in W.R.~Gilks S.~Richardson D.~S.,  ed., , Markov
  Chain Monte Carlo in Practice.
Chapman \& Hall, London

\bibitem[\protect\citeauthoryear{Fr{\"u}hwirth-Schnatter}{Fr{\"u}hwirth-Schnatter}{2006}]{Fruehwirth06}
Fr{\"u}hwirth-Schnatter S.,  2006, Finite Mixture and Markov Switching Models.
Springer Series in Statistics, Springer New York, \url
  {http://www.springer.com/us/book/9780387329093}

\bibitem[\protect\citeauthoryear{{Goulding} et~al.,}{{Goulding}
  et~al.}{2016}]{Goulding16}
{Goulding} A.~D.,  et~al., 2016, \mn@doi [\apj] {10.3847/0004-637X/826/2/167},
  \href {http://adsabs.harvard.edu/abs/2016ApJ...826..167G} {826, 167}

\bibitem[\protect\citeauthoryear{Lee \& Scott}{Lee \& Scott}{2012}]{Lee12.1}
Lee G.,  Scott C.,  2012, \mn@doi [Comput. Stat. Data Anal.]
  {10.1016/j.csda.2012.03.003}, 56, 2816

\bibitem[\protect\citeauthoryear{{Leistedt} et~al.,}{{Leistedt}
  et~al.}{2016}]{Leistedt16.1}
{Leistedt} B.,  et~al., 2016, \mn@doi [\apjs] {10.3847/0067-0049/226/2/24},
  \href {http://adsabs.harvard.edu/abs/2016ApJS..226...24L} {226, 24}

\bibitem[\protect\citeauthoryear{{Manjunath} \& {Wilhelm}}{{Manjunath} \&
  {Wilhelm}}{2012}]{Manjunath2012}
{Manjunath} B.~G.,  {Wilhelm} S.,  2012, preprint, \href
  {http://adsabs.harvard.edu/abs/2012arXiv1206.5387G} {} (\mn@eprint {arXiv}
  {1206.5387})

\bibitem[\protect\citeauthoryear{{McLachlan} \& {Jones}}{{McLachlan} \&
  {Jones}}{1988}]{McLachlan88.1}
{McLachlan} G.~J.,  {Jones} P.~N.,  1988, \mn@doi [Biometrics]
  {10.2307/2531869}, 44, 571

\bibitem[\protect\citeauthoryear{McLachlan \& Peel}{McLachlan \&
  Peel}{2000}]{McLachlan00.1}
McLachlan G.,  Peel D.,  2000, Finite Mixture Models, Wiley Series in
  Probability and Statistics.
John Wiley \& Sons, New York, \url
  {https://books.google.com/books?id=YXqflwEACAAJ}

\bibitem[\protect\citeauthoryear{Mengersen, Robert  \& Titterington}{Mengersen
  et~al.}{2011}]{Mengersen11.1}
Mengersen K.~L.,  Robert C.,   Titterington M.,  2011, Mixtures: estimation and
  applications.
 Vol. 896, John Wiley \& Sons

\bibitem[\protect\citeauthoryear{{Naim} \& {Gildea}}{{Naim} \&
  {Gildea}}{2012}]{Naim12.1}
{Naim} I.,  {Gildea} D.,  2012, preprint, \href
  {http://adsabs.harvard.edu/abs/2012arXiv1206.6427N} {} (\mn@eprint {arXiv}
  {1206.6427})

\bibitem[\protect\citeauthoryear{Nielsen}{Nielsen}{2000}]{Nielsen00}
Nielsen S.~F.,  2000, \mn@doi [Bernoulli] {10.2307/3318671}, 6, 457

\bibitem[\protect\citeauthoryear{{Rubin}}{{Rubin}}{1976}]{Rubin76.1}
{Rubin} D.~B.,  1976, \mn@doi [Biometrika] {10.1093/biomet/63.3.581}, 63, 581

\bibitem[\protect\citeauthoryear{Rubin}{Rubin}{1987}]{Rubin87.1}
Rubin D.,  1987, Multiple Imputation for Nonresponse in Surveys, Wiley Series
  in Probability and Statistics.
Wiley Series in Probability and Statistics, Wiley, \url
  {https://books.google.com/books?id=0KruAAAAMAAJ}

\bibitem[\protect\citeauthoryear{{Schafer} \& {Graham}}{{Schafer} \&
  {Graham}}{2002}]{Schafer02.1}
{Schafer} J.~L.,  {Graham} J.~W.,  2002, \mn@doi [Psychological Methods]
  {10.1037//1082-989X.7.2.147}, 7, 147

\bibitem[\protect\citeauthoryear{{Smyth} \& {Wolpert}}{{Smyth} \&
  {Wolpert}}{1999}]{Smyth99.1}
{Smyth} P.,  {Wolpert} D.,  1999, \mn@doi [Machine Learning]
  {10.1023/A:1007511322260}, 36, 59

\bibitem[\protect\citeauthoryear{Titterington, Smith  \& Makov}{Titterington
  et~al.}{1985}]{Titterington85.1}
Titterington D.,  Smith A.,   Makov U.,  1985, Statistical analysis of finite
  mixture distributions.
Wiley series in probability and mathematical statistics: Applied probability
  and statistics, Wiley, \url {https://books.google.com/books?id=hZ0QAQAAIAAJ}

\bibitem[\protect\citeauthoryear{Ueda, Nakano, Ghahramani  \& Hinton}{Ueda
  et~al.}{2000}]{Ueda00.1}
Ueda N.,  Nakano R.,  Ghahramani Z.,   Hinton G.~E.,  2000, \mn@doi [Journal of
  VLSI signal processing systems for signal, image and video technology]
  {10.1023/A:1008155703044}, 26, 133

\bibitem[\protect\citeauthoryear{{Wolpert}}{{Wolpert}}{1992}]{Wolpert92.1}
{Wolpert} D.~H.,  1992, \mn@doi [Neural Netw.] {10.1016/S0893-6080(05)80023-1},
  5, 241

\bibitem[\protect\citeauthoryear{Wolynetz}{Wolynetz}{1979}]{Wolynetz79.1}
Wolynetz M.~S.,  1979, Journal of the Royal Statistical Society. Series C
  (Applied Statistics), 28, 195

\bibitem[\protect\citeauthoryear{Wu}{Wu}{1983}]{Wu83.1}
Wu C. F.~J.,  1983, The Annals of Statistics, 11, 95

\bibitem[\protect\citeauthoryear{Zhang, Chen, Sun  \& Chan}{Zhang
  et~al.}{2003}]{Zhang03.1}
Zhang Z.,  Chen C.,  Sun J.,   Chan K.~L.,  2003, \mn@doi [Pattern Recognition]
  {http://dx.doi.org/10.1016/S0031-3203(03)00059-1}, 36, 1973

\makeatother
\end{thebibliography}

\end{document}